\documentclass[10pt,journal,compsoc]{IEEEtran}
\usepackage[latin9]{inputenc}
\usepackage{array}
\usepackage{multirow}
\usepackage{amstext}
\usepackage{graphicx}
\usepackage{subfig}
\usepackage{algorithm}
\usepackage{adjustbox}
\usepackage{tabulary}
\usepackage{float}
\usepackage[noend]{algpseudocode}

\usepackage{amsmath}
\usepackage[nocompress]{cite}

\ifCLASSOPTIONcompsoc
  \usepackage[nocompress]{cite}
\else
  \usepackage{cite}
\fi
\ifCLASSINFOpdf
\else
\fi
\begin{document}
%
\title{A Directed Acyclic Graph Approach \\ to Online Log Parsing}

\author{Pinjia~He,
        Jieming~Zhu,
        Pengcheng Xu,
        Zibin Zheng,~\IEEEmembership{Senior Member,~IEEE,}
        and~Michael R.~Lyu,~\IEEEmembership{Fellow,~IEEE}
\IEEEcompsocitemizethanks{\IEEEcompsocthanksitem P. He, P. Xu and M. R. Lyu are with the Department of Computer Sciences and Engineering, The Chinese University of Hong Kong, Shatin, NT, Hong Kong. E-mail: \{pjhe, lyu\}@cse.cuhk.edu.hk, xu\_pengcheng@link.cuhk.edu.hk.\protect\\
\IEEEcompsocthanksitem J. Zhu is with Huawei 2012 Labs, Huawei, Shenzhen, China. Email: jmzhu@ieee.org\protect\\
\IEEEcompsocthanksitem Z. Zheng is with the School of Data and Computer Science, Sun Yat-sen University, Guangzhou, Guangdong, China. Email: zhzibin@mail.sysu.edu.cn\protect\\
}
\thanks{Manuscript received XX, XX; revised XX, XX.}}

\markboth{Journal of \LaTeX\ Class Files,~Vol.~X, No.~X, X~X}%
{Shell \MakeLowercase{\textit{et al.}}: tpds}

\IEEEtitleabstractindextext{%
\begin{abstract}
Logs are widely used in modern software system management because they are often the only data accessible that record system events at runtime. In recent years, because of the ever-increasing log size, data mining techniques are often utilized to help developers and operators conduct system reliability management. A typical log-based system reliability management procedure is to first parse log messages because of their unstructured format; and apply data mining techniques on the parsed logs to obtain critical system behavior information. Most of existing research studies focus on offline log parsing, which need to parse logs in batch mode. However, software systems, especially distributed systems, require online monitoring and maintenance. Thus, a log parser that can parse log messages in a streaming manner is highly in demand. To address this problem, we propose an online log parsing method, namely Drain, based on directed acyclic graph, which encodes specially designed rules for parsing. Drain can automatically generate a directed acyclic graph for a new system and update the graph according to the incoming log messages. Besides, Drain frees developers from the burden of parameter tuning by allowing them use Drain with no pre-defined parameters. To evaluate the performance of Drain, we collect 11 log datasets generated by real-world systems, ranging from distributed systems, Web applications, supercomputers, operating systems, to standalone software. The experimental results show that Drain has the highest accuracy on all 11 datasets. Moreover, Drain obtains 37.15\%$\sim$ 97.14\% improvement in the running time over the state-of-the-art online parsers. We also conduct a case study on a log-based anomaly detection task using Drain in the parsing step, which determines its effectiveness in system reliability management. 
\end{abstract}

}

\maketitle

\IEEEdisplaynontitleabstractindextext

%
\IEEEpeerreviewmaketitle


\IEEEraisesectionheading{\section{Introduction}}\label{sec:introduction}

\IEEEPARstart{M}{}odern software systems, such as search engines, cloud services, and online chatting, play an indispensable role in our daily life. Most of these systems run in a distributed manner. Different from traditional systems, modern software systems often adopt the 24/7 operation, because the users of these distributed systems expect available services at any time. Any non-trivial downtime of these systems can cause enormous revenue loss \cite{downtimeDatacenter, downtimeFacebook, downtimeAmazon} to both service providers (e.g., Amazon EC2 \cite{amazonec2} and Microsoft Azure \cite{microsoftazure}) and service users (e.g., Trello \cite{trello} and Slack \cite{slack}). Thus, the reliability assurance of software systems is of considerable importance. 

Logs, which record system runtime information, are widely used in various reliability assurance tasks. Typical log-based reliability assurance tasks include anomaly detection \cite{weixu09, qfu09, helenGuSRDS11, Fu12SRDS}, fault diagnosis \cite{wong2012effective, zou2016uilog}, program verification \cite{dingATC15, wshang13}, and performance monitoring \cite{Nagaraj12, dingATC15}. In general, each log message is a line of text printed by logging statements (e.g., \textit{printf(), logging.info()}) written by developers. With respect to system runtime, each logging statement generates log messages of the same log event (i.e., log template), which describes a specific system operation.  In recent years, as distributed systems have become larger in scale and more complex in structure, the volume of logs has increased rapidly (e.g., 50 GB/h \cite{miTPDS13}). Thus, traditional log-based methods that mainly rely on manual inspection and analysis have become impractical and prohibitive. To accelerate the log analysis process, a large number of data mining models (e.g., classification models) have been adopted by both researchers and practitioners. 

Although these data mining models facilitate automated log analysis techniques, most of them require structured data as the input (e.g., a list of log events or a matrix). However, raw log messages are usually unstructured, because developers are allowed to write free-text log messages in the source code, and typical logging statements also record program variables for post-mortem analysis. Thus, to obtain structured input data, log parsers have been used to transform unstructured log messages into structured log events. An unstructured log message, as in the following example, usually contains a set of fields that record system runtime information: \textit{timestamp} (records the occurrence time of an event), \textit{verbosity level} (indicates the severity level of an event, e.g., INFO), and \textit{raw message content} (free-text description of a system operation).

\begin{footnotesize} \begin{verbatim}
081109 204655 556 INFO dfs.DataNode$PacketResponder
: Received block blk_3587508140051953248 of size 67
108864 from /10.251.42.84
\end{verbatim} \end{footnotesize}

Traditional log parsing involves heavy use of regular expressions \cite{SEC13}. In particular, for each log event, developers manually construct a regular expression and add it to a list of parsing rules. With respect to system runtime, to parse system logs, a traditional log parser compares an incoming log message with all of the regular expressions in the list and returns the matched regular expression as the log event for the log message. However, traditional methods have become impractical and error-prone because of the following reasons. First, as the systems have become larger in scale, the volume of logs has increased rapidly. Second, modern systems have become complex in structure. The source codes of a system can be written by hundreds of developers all over the world, because developers tend to use open-source components online (e.g., Github) or directly invoke the existing Web services. Thus, developers who maintain the parsing rules may have no idea of the original logging purpose, which further increases the difficulty of manual log parsing. Third, because of the wide adoption of agile software development, the logging statements of modern systems update frequently (e.g., hundreds of new logging statements every month \cite{weixutheis}). Thus, developers need to periodically update the parsing rules to guarantee parsing accuracy for a new version.

Recent studies have proposed various automated log parsing methods to ease the burden of developers. Xu et al. \cite{weixu09} mined parsing rules from source codes using  program static analysis techniques. Although effective, they require program source codes as the input, which are often inaccessible (e.g., third-party libraries and Web services). In this study, we focus on data-driven log parsers that only require system logs as input. Most of the existing data-driven log parsers \cite{SLCT03, ltang11, qfu09, IPLoM12, HeTDSC17} work in an offline manner. In practice, developers use these log parsers on existing system logs to automatically mine parsing rules, and parse the incoming log messages on the basis of these parsing rules. However, to keep the parsing rules fresh, developers need to regularly re-run the parser to spot possible updates of logging statements. Moreover, modern systems often collect system logs in a streaming manner. For example, a typical log collection system has a log shipper installed on each node to forward the log messages in a streaming manner to a centralized server that contains a log parser \cite{spellICDM16}. Thus, an online log parser that can parse log messages online and dynamically update the parsing rules is in high demand. There are a few preliminary studies on online log parsing \cite{Mizutani13, spellICDM16}. However, we have observed that these parsers are not sufficiently accurate and efficient, which makes them ineligible for log parsing in modern systems. Furthermore, the existing methods require a considerable amount of parameter tuning work, which leads to unnecessary manual effort for the developers. Moreover, the pre-defined parameters might limit the robustness of the online parsers against the logging statement updates.

To address these problems, we propose an online log parsing method, called Drain; it can parse log messages in a streaming manner and update its parsing rules during runtime. In particular, we develop a directed acyclic graph (DAG), which encodes specially designed heuristic rules. To parse the incoming log messages, Drain uses a DAG to separate them into disjoint log groups, where the log messages in a log group have the same log event. The depth of a DAG is fixed to accelerate the parsing process. Furthermore, different from the existing log parsers, the core part of Drain does not require any pre-defined parameters from the developers. The parameters used are initialized automatically and updated dynamically according to the incoming log messages. 

To evaluate the performance of Drain, we collect 11 log datasets generated by real-world systems, ranging from distributed systems, Web applications, supercomputers, operating systems, to standalone software. To the best of our knowledge, this study is the first work that evaluates log parsers on such amount and variety of datasets. We compare Drain with three offline log parsers and two online log parsers in terms of accuracy and efficiency. The experimental results demonstrate that Drain exhibits the highest parsing accuracy on all of the log datasets, and achieves 37.15\%$\sim$ 97.14\% improvement in the running time over the state-of-the-art online log parser.

In summary, this study makes the following contributions:
\begin{itemize}
  \item It presents an online log parsing method (Drain) based on a DAG, which parses system logs in a streaming manner and can automatically adapt to system updates during runtime.
  \item Drain can initialize its parameters automatically and update them dynamically according to the incoming logs.
  \item Extensive experiments have been conducted on 11 log datasets of real-world systems to evaluate the performance of Drain in terms of the accuracy, efficiency, and effectiveness in system reliability management. 
  \item This work releases the first set of log parsing data \cite{logparserdatasets}, including logs from distributed systems, Web applications, supercomputers, operating systems, and standalone software. The source code of Drain \cite{cuhklogparser} has also been released for reproducible research.
\end{itemize}

Extended from its preliminary conference version \cite{He17ICWS}, the paper makes several major enhancements: the design and implementation of a new log parsing method on the basis of DAG; automated parameter tuning mechanism; experimental comparison on one more log parser and six more datasets; a discussion highlighting the potential limitations and future work; the release of 11 datasets for log parsing \cite{logparserdatasets}; and code release of Drain for reproducible research.

The remainder of this paper is organized as follows. Section \ref{sec:overview} presents an overview of log parsing. Section \ref{sec:methodology} describes our online log parsing method, Drain, with the focus on the structure of DAG, while Section \ref{sec:update} introduces the dynamic update of DAG. We introduce the released datasets in Section \ref{sec:dataset}. Then we present the experimental results in Section \ref{sec:exper}.  We discuss potential limitations and future work in Section \ref{sec:dis}. Related work is introduced in Section \ref{sec:related}. Finally, we conclude this paper in Section \ref{sec:con}.

\section{Overview of Log Parsing}\label{sec:overview}

\begin{figure}
\centering{}\includegraphics[scale=0.56]{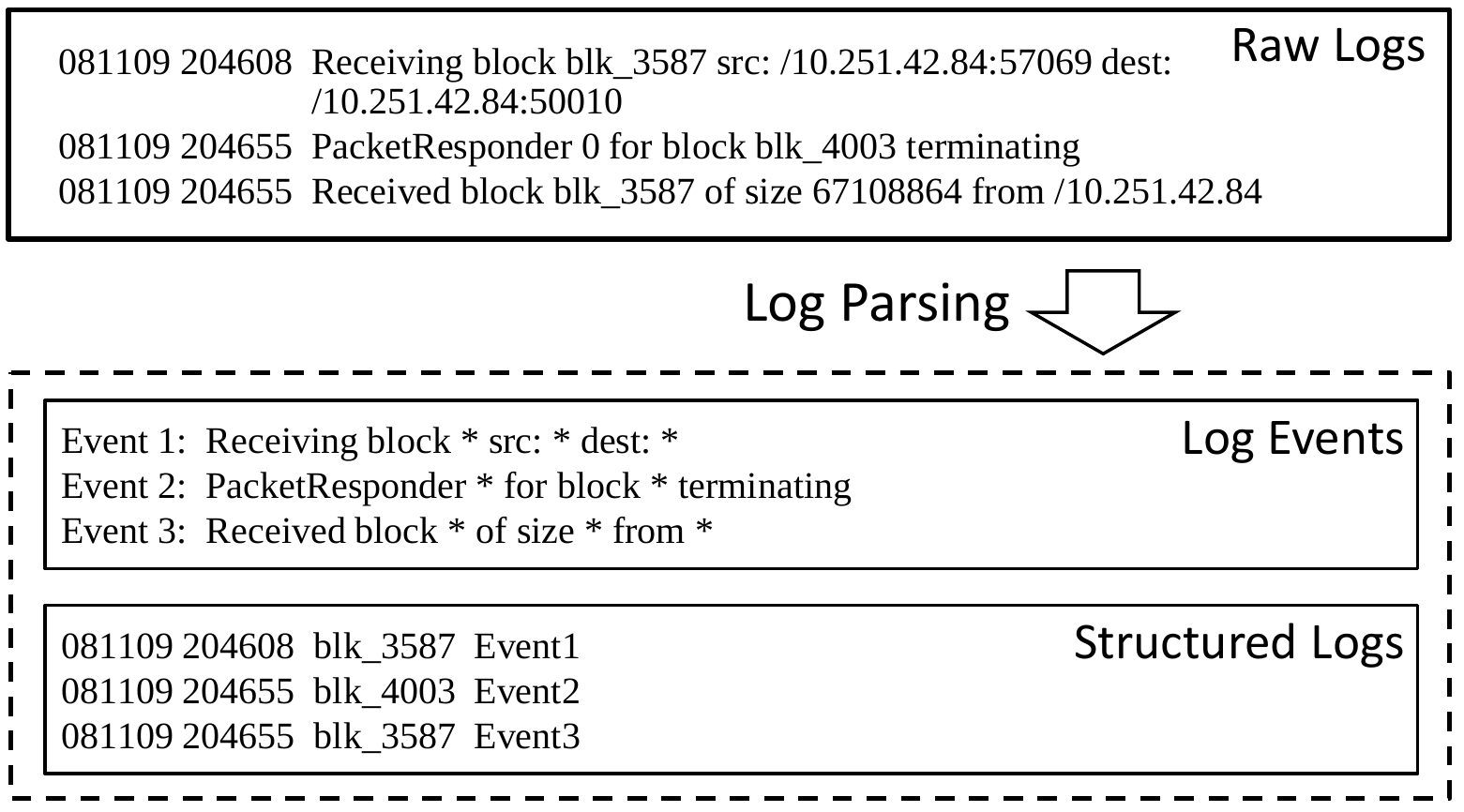}\protect\caption{Overview of Log Parsing}
\label{fig:rawlogmessage}
\end{figure}

Log parsing is a process that transforms unstructured log messages into structured log messages. In particular, a log parser matches each log message with a log event. Thus, a log parser needs to distinguish between the \textit{constants} and \textit{variables} in each unstructured log message. Fig. \ref{fig:rawlogmessage} illustrates an example of log parsing, wherein log messages are simplified Hadoop Distributed File System (HDFS) raw log messages collected on the Amazon EC2 platform \cite{weixu09}. ``Receiving" in the first raw log message is a constant, while ``blk\_3587" represents a variable. All of the constants in a log message form the corresponding log event, which describes a specific system operation, for example, ``Receiving block src: dest:". In practice, we often replace the variables with asterisks to make the log event more informative.

Log parsing can be regarded as a clustering process, whereby log messages with different log events are clustered into different log groups. Then, developers can use ad-hoc scripts to generate structured logs. In this study, we also model log parsing as a clustering process, which aligns with typical log parsers proposed by existing papers \cite{SLCT03,ltang11,qfu09,IPLoM12,HeTDSC17, Mizutani13, spellICDM16, He17ICWS}. As illustrated in Fig. \ref{fig:rawlogmessage}, the input of Drain is the raw logs generated during system runtime, while the output contains two parts: log events and structured logs. In particular, log events are templates mined from the incoming raw logs, which show all of the triggered system operations. Structured logs contain the event ID and the fields of interest (e.g., timestamp and block ID). Based on the structured logs, developers can easily apply various automated log analysis techniques. 


\section{Methodology}\label{sec:methodology}

\begin{figure*}[t]
\centering{}
\includegraphics[scale=0.55]{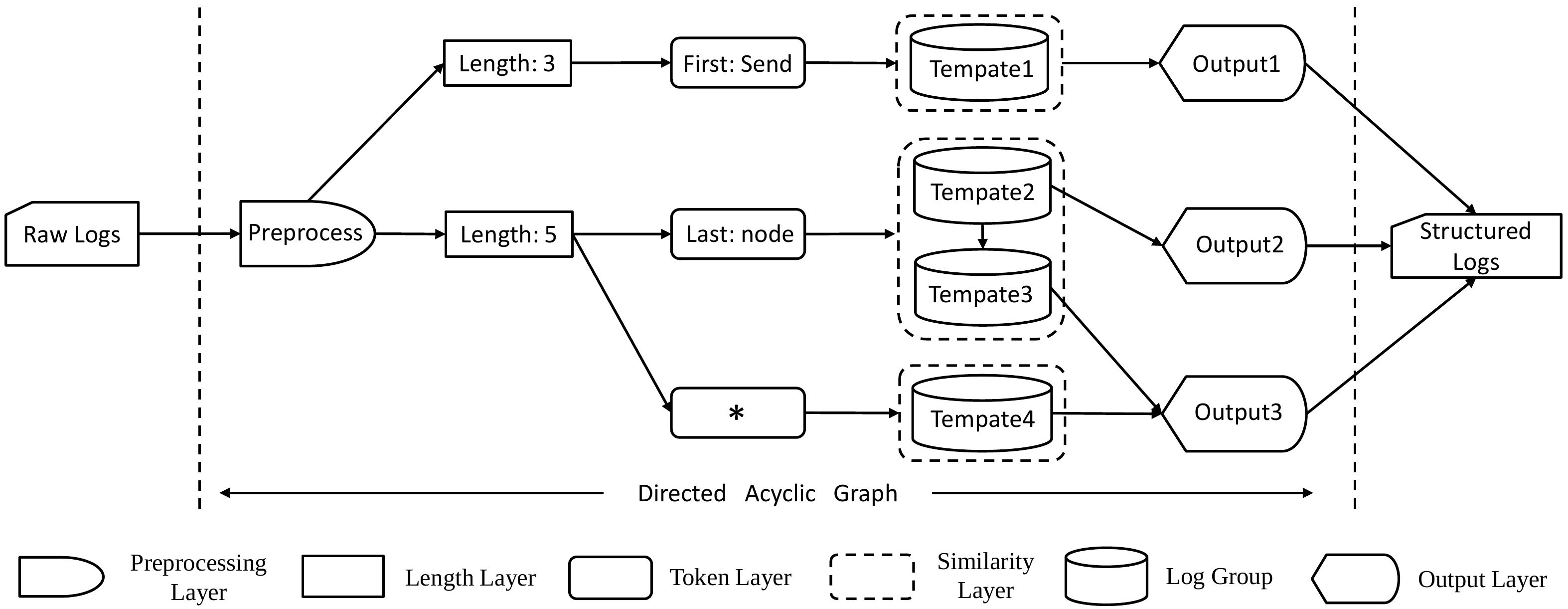}
\caption{Example of Directed Acyclic Graph Generated by Drain}
\label{fig:dag}
\end{figure*}

In this section, we introduce the proposed online log parsing method, called Drain, which is based on a DAG. As illustrated in Fig. \ref{fig:dag}, the DAG used by Drain has the following five layers: preprocessing layer, length layer, token layer, similarity layer, and output layer. We will introduce the details of these layers in the following subsections.

\subsection{Overview of DAG}

A real-world system generates tons of logs every day, and the newly generated logs arrive in a streaming manner. To parse a new raw log message, an online log parser matches the log message with the corresponding log event. In this process, a simple solution is to compare the raw log message with all of the candidate log events and return the log event that has the most similarity. However, this solution is very slow because the number of log event types increases rapidly in parsing. To make this search process more efficient, Drain uses a DAG.

For each system, Drain automatically generates a DAG during runtime. As illustrated in Fig. \ref{fig:dag}, a DAG has five node types. The nodes with the same node type form a layer. When a raw log message arrives, it is preprocessed on the basis of the domain knowledge in the preprocessing layer. Then, Drain traverses along a path according to the characteristics (e.g., number of tokens in the log message) of the log message. At the end, Drain traverses to a node in the output layer, which matches the raw log message with a log event, and outputs a structured log message.

In the following text, we will introduce the design of nodes in the five layers of the DAG. Note that unless otherwise stated, we use the DAG shown in Fig. \ref{fig:dag} as an example in the following explanation.

\subsection{Preprocessing Layer}
In the first layer of DAG, we preprocess the raw log message when it arrives. Specifically, Drain allows users to provide simple regular expressions based on domain knowledge that represent commonly-used variables, such as IP address and block ID. When Drain matches a token by a regular expression in the raw log message, it will replace the token matched by a user-defined constant. For example, block IDs in Fig. \ref{fig:rawlogmessage} will be matched by regular expression ``blk\_[0-9]+" in the preprocessing layer, and further be replaced by ``blkID". Although some offline parsers (e.g., \cite{qfu09, ltang11}) have the preprocessing step, it has not been used by existing online log parsers.

Intuitively, preprocessing is effective because it replaces variables (e.g., blk\_3587) in the log messages by constants (e.g., blkID), which provides additional information to the log parser. Note that the regular expressions used in the preprocessing layer are different from those used by traditional log parsing methods. Specifically, the preprocessing layer aims at pruning well-known variables. Thus developers only need to come up with simple regular expressions for variables with common patterns. In traditional methods, however, developers need to construct a regular expression to match the whole log message. A regular expression for the whole log message is much more complex than that for a variable, therefore requiring more human effort. Moreover, in traditional methods, developers need to construct a list of regular expressions to match the whole log dataset; while in Drain's preprocessing layer, developers only need to construct several simple regular expressions. For example, for log dataset BGL, developers need to construct 376 regular expressions to match all the logs by using traditional methods. If developers use Drain, they only need to specify one simple regular expression (i.e., ``core\.[0-9]+" for core ID). In this paper, the datasets used in our evaluation section require at most two such regular expressions.

\subsection{Length Layer}
After preprocessing, the raw log message will traverse to a node in the length layer. Nodes in the length layer separate preprocessed log messages into different paths, where each length layer node represents log messages of a specific log message length. In this paper, by log message length, we mean the number of tokens in a log message. For example, the log message ``Send file file\_01" contains three tokens, and thus its log message length is three. After preprocessing, this log message will traverse to the ``Length: 3" node in Fig. \ref{fig:rawlogmessage} guided by Drain.

The length layer is based on the assumption that log messages with the same log event will be more likely to have the same log message length. Although it is possible that log messages with the same log event have different log message lengths, they can be handled by simple postprocessing. To address these relatively rare cases, Drain provides an optional mechanism to merge similar log messages of different log message lengths in the output layer, which will be introduced later. 


\subsection{Token Layer}\label{sec:tokenlayer}
Nodes in the token layer separate log messages according to their \textit{split token}, which is a token that guides the path traversing. Drain considers three conditions when finding the split token for a log message as follows. (1) The first token of the log message is the split token. (2) The last token of the log message is the split token. (3) The log message does not contain a split token.

\begin{algorithm}[t]
\caption{
Drain Token Layer: Split token selection.  
}
\label{alg:tokenlayer}
\begin{algorithmic}[1]
\Require a log message from a length node: $logMsg$

\Ensure a split token: $splitToken$ 

\State $firstToken\leftarrow$ first token in $logMsg$ 
\State $lastToken\leftarrow$ last token in $logMsg$ 
\State $splitToken\leftarrow None$ \Comment{Initialization of the split token}

\If{\textit{\Call{hasDigit}{$firstToken$}$=true$}}

    \If{\textit{\Call{hasDigit}{$lastToken$}$=false$}}
    	\State $splitToken\leftarrow lastToken$
    \EndIf

\Else

    \If{\textit{\Call{hasDigit}{$lastToken$}$=true$}}
    	\State $splitToken\leftarrow firstToken$
    \Else
    	\If{\textit{\Call{hasPun}{$firstToken$}$=true$}}
    		\If{\textit{\Call{hasPun}{$lastToken$}$=false$}}
    			\State $splitToken\leftarrow lastToken$
    		\EndIf
    	\Else
    		\State $splitToken\leftarrow firstToken$
    	\EndIf
    \EndIf

\EndIf    

\Statex 
\Function{hasDigit}{$token$}
	\ForAll{\textit{$c$ in $token$}} \Comment{$c$ is a character}
		\If{\textit{$c$ is a digit}}
			\State \Return $true$
		\EndIf
	\EndFor
	\State \Return $false$
\EndFunction

\Statex 
\Function{hasPun}{$token$}
	\ForAll{\textit{$c$ in $token$}} \Comment{$c$ is a character}
		\If{\textit{$c$ is a special character}}
			\State \Return $true$
		\EndIf
	\EndFor
	\State \Return $false$
\EndFunction
\end{algorithmic}
\end{algorithm}

Algorithm \ref{alg:tokenlayer} provides the pseudo code of the split token selection process in the token layer. In this step, Drain selects a token in the log message as the split token, which will be further used to guide the log message traversing process. For example, if Drain selects the first token of ``Send file file\_01" as the split token, Drain will inspect its first token ``Send". Based on the first token, the log message will traverse to the ``First: Send" node in the token layer, which is pointed by the ``Length: 3" node in the length layer. Specifically, Algorithm \ref{alg:tokenlayer} checks whether the first token or the last token contains any digits or special characters. At the beginning, both the first token and the last token are regarded as potential split tokens, and the split token is initialized by ``None", which means currently the log message has no split token (lines 1$\scriptsize{\sim}$3). We focus on the first token and the last token in this layer, because according to our previous experience on log parsing \cite{He16DSN, He17ICWS, HeTDSC17}, developers tend to either start or end a log message with a constant. For example, log message ``Send file file\_01" starts with the constant ``Send", while log message ``10 bytes are sent" ends with the constant ``sent". 

However, not all first tokens or last tokens are regarded as the split tokens. Drain considers tokens containing digits are relatively unlikely to be constants, such as block IDs (e.g., blk\_3587). Thus, if the first (last) token contains digits while the last (first) token does not, Drain selects the last (first) token as the split token, which is line 4$\scriptsize{\sim}$6 (line 7$\scriptsize{\sim}$9) in Algorithm \ref{alg:tokenlayer}. If neither of the two tokens contain digits, Drain will further check whether they contain special characters (line 11$\scriptsize{\sim}$15). If the first token contains special characters while the last token does not, Drain selects the last token as the split token (line 11$\scriptsize{\sim}$13). Otherwise, the first token is selected as the split token (line 14$\scriptsize{\sim}$15). Besides, if both the first token and the last token do not contain any digits or special characters, the first token will be selected as the split token (line 14$\scriptsize{\sim}$15). The special characters used in this paper for all datasets are string punctuations such as ``\textit{\#\^\$'*+,/$<$=$>$@\_`)$|$$\sim$}", which are common special characters in system logs. Besides, to make Drain more flexible in practice, we allow developers to set customized special characters. If both the first token and the last token contain digits, Drain considers the current log message does not have a split token, and the $splitToken$ in Algorithm \ref{alg:tokenlayer} remains $None$. In this case, it will traverse to the ``*" node in the token layer. 


\subsection{Similarity Layer}

Log messages that arrive at the same similarity node are of the same log message length and contain the same split token. To further split log messages into different log groups, for each similarity node, Drain maintains a list of log groups. Each log group contains a log event, which is the template mined by Drain to represent the log group, and a list of log IDs. A log event contains tokens and wildcards. Tokens are potential constants, while wildcards are mined variables. For example, ``Send file *" could be a log event of a log group. ``Send" and ``file" are potential constants, while ``*" is a variable.

In the similarity layer, for the incoming log message, Drain will choose the most suitable log group in the similarity node, and match it with the corresponding log event. Specifically, Drain selects the most suitable log group according to the similarity between the incoming log message and the log events in the log groups. To calculate the similarity between the two texts, we define $simSeq$ as follows:

\begin{equation}
\label{equ:simseq}
simSeq=\frac{\sum_{i=1}^{n} {equ(seq_1(i), seq_2(i))}}{n_c},
\end{equation}

\hspace{-3.2ex}where $seq_1$ and $seq_2$ represent the log message and the log event respectively; $seq(i)$ is the $i$-th token of the sequence; $n$ is the log message length of the sequences; $n_c$ is the number of tokens in the log event that are not variables (i.e., ``*"); and function $equ$ is defined as following:

\begin{equation}
\label{equ:equ}
 equ(t_1, t_2) = 
  \begin{cases} 
   1 & \text{if } t_1 = t_2 \text{ and } t_2 \neq *\\
   0       & \text{otherwise }
  \end{cases}
\end{equation}

\hspace{-3.2ex}where $t_1$ and $t_2$ are two tokens. Intuitively, in Equation \ref{equ:simseq}, we conduct token-wise comparison between the log message and the log event. If the token in the log event is ``*", we skip to the next token pairs. In the log events mined by Drain, ``*" represents a variable. To calculate the similarity between two sequences, we focus on their constants. Thus, we only count when the two compared tokens are the same and the tokens from the log event are not variables.

We calculate $simSeq$ for each log group in the current similarity node, and select the log group with the highest $simSeq$. If several log groups present the same $simSeq$, we select the log group whose log event contains the fewest variables. After log group selection, we compare it with a similarity threshold $st$. If $simSeq \geq st$, Drain matches the selected log group with the incoming log message and proceeds to the next layer. Otherwise, Drain will report that it encounters a log message whose log event has not been recorded yet. In this case, Drain will create a new log group and update the DAG accordingly. After the update process, the log message will traverse to the next layer from the newly generated log group. The update process will be introduced in detail in Section \ref{sec:update}.

\subsection{Output Layer}
\begin{figure}[t]
\centering{}
\includegraphics[scale=1.1]{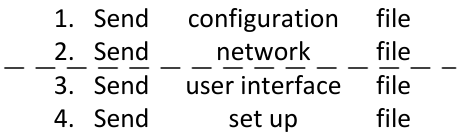}
\caption{An Example of Over-parsing}
\label{fig:outputlayer}
\end{figure}

At this point, different log messages have arrived at different log groups in the similarity layer. Thus, before traversing to the output layer,  a log message has already matched with a log event. Intuitively, a log parser can already output a structured log when the log message matches a log event. However, in practice, developers may want to merge some over-parsed log groups. For example, Fig. \ref{fig:outputlayer} presents four log messages, which have the same log event ``Send * file". However, these four log messages will traverse to different paths in Drain's DAG in the length layer because of their different log message length. Although the over-parsing condition is relatively rare, we provide an optional merge mechanism in the output layer of Drain to merge these over-parsed log groups. If two log groups are merged, they will point to the same output layer node (e.g., group ``Template3" and ``Template4" in Fig. \ref{fig:dag}).

In the search process, the log message will traverse to the connected node in the output layer. If its current log group merges with other log groups in the output layer, Drain will update the log template accordingly. Finally, Drain outputs the structured log message.

\subsection{Cache Mechanism}

We have introduced the search process of Drain that searches a path in the DAG to match a log message with a log event. Essentially, DAG accelerates the parsing process by setting search rules in different layers (e.g., the length layer nodes). Thus, a log message can find the most suitable log group by comparing with a subset of existing log events instead of all existing log events. 

To further accelerate the search process, we design a cache mechanism for Drain. Specifically, for each node in the length layer, Drain maintains a pointer from the node to a log group in a similarity layer node. After parsing a log message, Drain will update the pointer from the corresponding length layer node to the log group matched by Drain. This cache mechanism is useful for modern systems that often generate a sequence of log messages of the same log event, such as some distributed systems or supercomputers. For example, nodes in supercomputers often generate log messages of the same log event simultaneously.

\begin{figure}[t]
\centering{}
\includegraphics[scale=0.45]{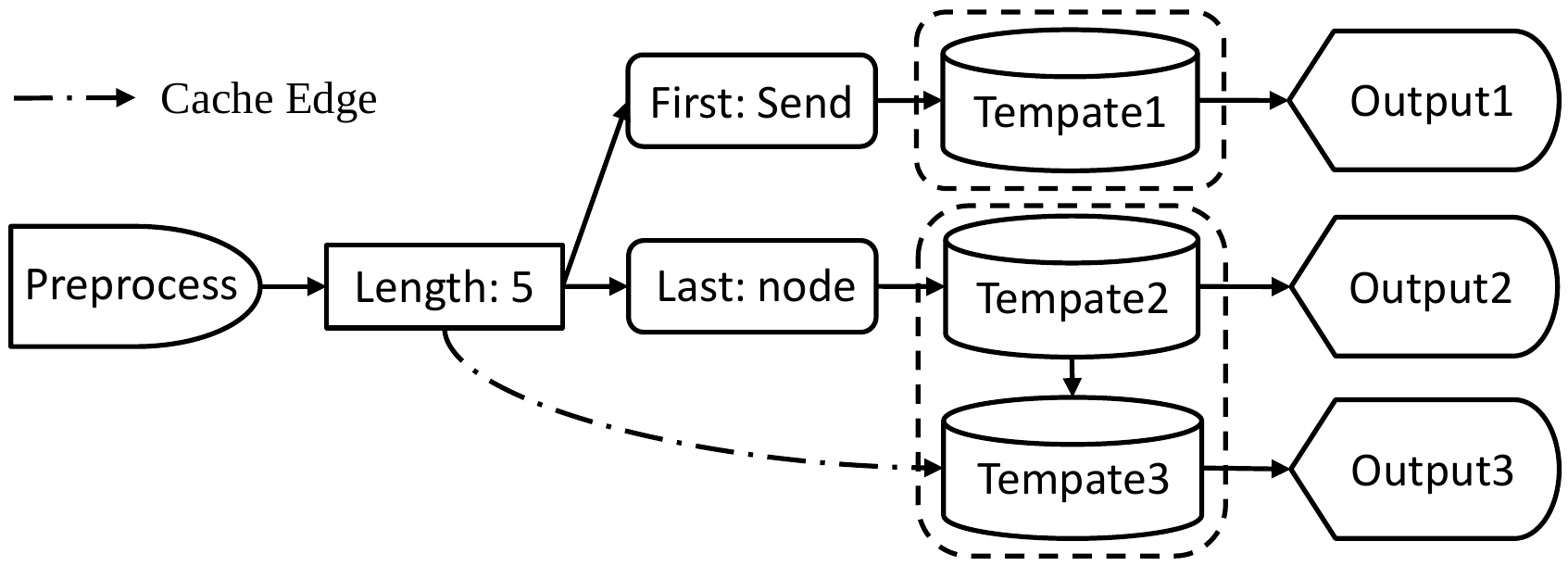}
\caption{Cache Example in DAG}
\label{fig:cache}
\end{figure}

With the cache mechanism, Drain fetches the cached log group first, and calculates the similarity between the incoming log message with the log event of the returned log group. If the similarity is larger than the similarity threshold $st$, Drain directly jumps to the cached log group and further proceeds to the next layer (i.e., output layer). Otherwise, Drain uses the normal search strategy introduced in the previous sections. As illustrated in Fig. \ref{fig:cache}, the cache mechanism maintains a cache edge from the length layer node ``Length: 5" to a specific log group in the similarity layer. With this cache edge, Drain can fetch the cached log group ``Template3" first for similarity computation.

\section{Update DAG}\label{sec:update}

In this section, we introduce the dynamic update process of Drain. After searching for the most suitable log group by searching the DAG, Drain updates the DAG automatically. In particular, if Drain cannot match the log message with an existing log group, it generates a new log group. Otherwise, Drain updates an existing log group. In both cases, Drain  (1) updates the DAG and (2) conducts automated parameter tuning. In the following, we will introduce these two cases in detail. Finally, we will introduce the optional merge mechanism that allows developers to merge the over-parsed log groups.

\subsection{Add a New Log Group}

\subsubsection{Update the DAG}
\begin{figure}[t]
\centering{}
\includegraphics[scale=0.45]{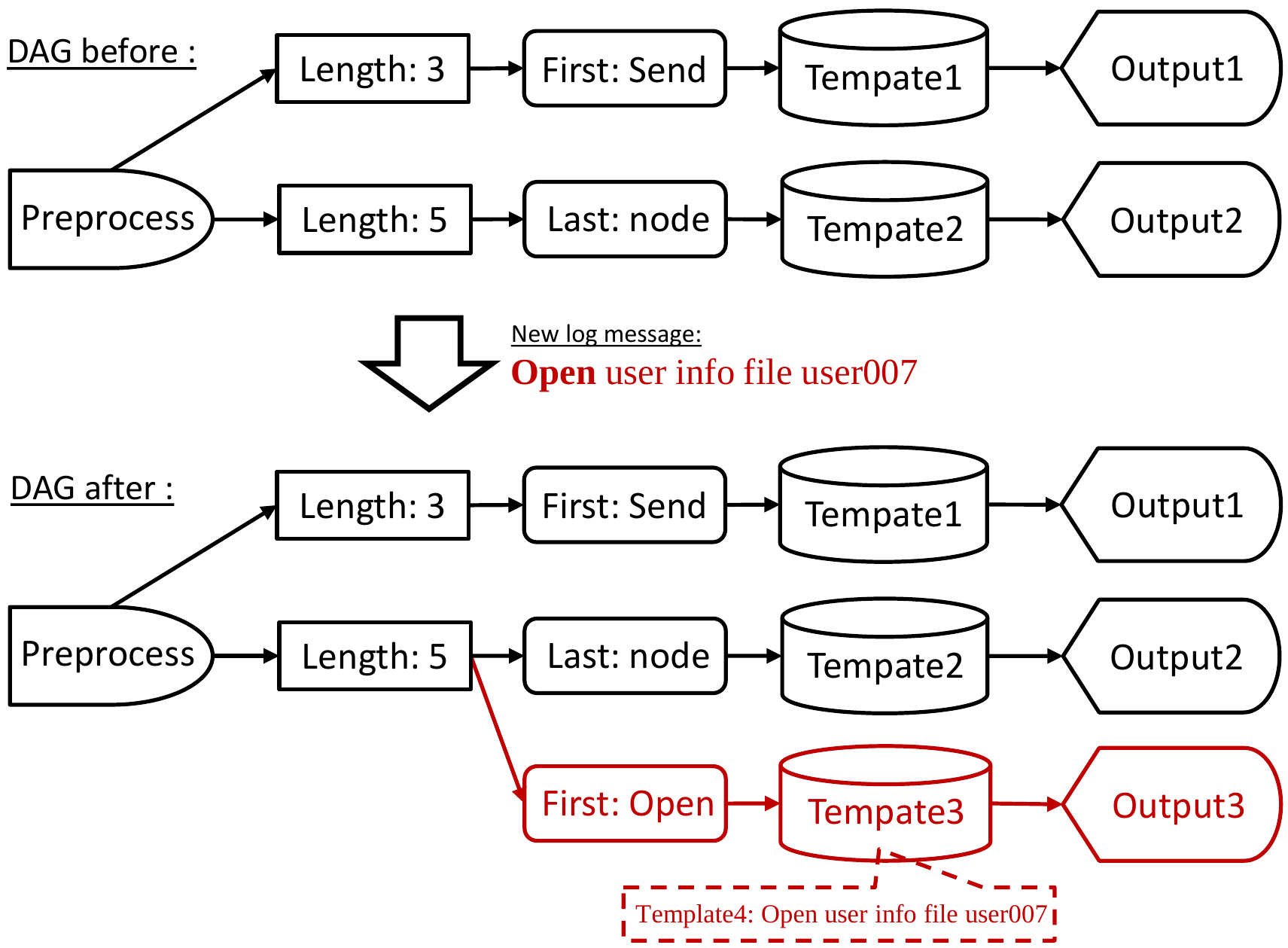}
\caption{DAG Update Example}
\label{fig:dagupdate}
\end{figure}

When Drain cannot match the current log message with any existing log groups, Drain will generate a new log group, where the log message itself is the corresponding log event. Then, Drain will add this log group to the DAG by traversing the DAG. For simplicity, we illustrate an example in Fig. \ref{fig:dagupdate} with an incoming log message ``Open user info user007".

Specifically, Drain calculates the log message length of the current log message, and checks whether there is a related node in the length layer of the DAG. For example, as illustrated in Fig. \ref{fig:dagupdate}, when the new log message arrives, Drain checks whether ``Length: 5" is already in the DAG. If yes, Drain traverses to node ``Length: 5". Otherwise, Drain creates a ``Length: 5" node and points the preprocessing node to the newly generated node. After that, Drain selects the split token using the strategy illustrated by Algorithm \ref{alg:tokenlayer}. In our example, the first token ``Open" will be selected as the split token. Then, Drain checks whether the current length layer node (i.e., ``Length: 5") points to a token layer node ``First: Open". If yes, Drain traverses to node ``First: Open". Otherwise, Drain creates a ``First: Open" node and points the ``Length: 5" node to the newly generated node. 

Then, if a token layer node is created, Drain will create a node in the similarity layer, which is a list containing the new log group as its first element. Then the new token layer node is pointed to the new similarity layer node. If Drain traverse to an existing token layer node, it will traverse to the similarity layer node. Note that a token layer node only points to one similarity layer node. After that, Drain appends the new log group to the list of the similarity layer node. In both cases, Drain creates a output layer node, and points the new log group to the new output layer node. By default, each log group points to one output layer node. However, Drain provides an optional merge mechanism that allows developer to merge the output layer nodes. This optional mechanism will be introduced at the end of this section. For simplicity, in Fig. \ref{fig:dagupdate}, we do not show the similarity layer node, since each simlarity layer node only contains one log group in the figure.

\subsubsection{Similarity Threshold Initialization}
For each log group, Drain provides a parameter $st$, which is employed as the similarity threshold in the search phase. Specifically, each node in the similarity layer contains a list of log groups. In the search phase, when the log message arrives a node in the similarity layer, Drain will choose the most suitable log group from the list based on similarity calculation in Equation \ref{equ:simseq}. If the similarity is larger than $st$, Drain will match the log message with the most similar log group.

Considering the similarity threshold ($st$ for Drain), existing online parsers (i.e., \cite{Mizutani13, spellICDM16}) provide parameters that require manual tuning for different system logs. Specifically, SHISO \cite{Mizutani13} provides four parameters, while Spell \cite{spellICDM16} requires one parameter. SHISO \cite{Mizutani13} uses $t_s$ as the similarity threshold to find the most suitable log group. This often causes unnecessary human effort, because developers need to tune the parameters required for each system they use. Besides, logging statements in modern systems update frequently (e.g., hundreds of new logging statement each month \cite{weixutheis}). Thus, parameters tuned could become ineffective as modern systems evolve, which leads to inaccuracy in log parsing. Moreover, we expect that an online log parser is self-adaptive to the update of logging statements and the change of log datasets. To address these problems, we propose self-adjusting similarity threshold $st$, which initializes itself automatically and updates dynamically. Different from existing parsers that the parameters are used globally, which means that different log groups have the same pre-defined similarity threshold, Drain assigns different similarity thresholds (i.e., $st$) to different log groups. Specifically, the initial value of $st$, namely $st_{init}$ is calculated as follows:

\begin{equation}
\label{equ:initst}
st_{init}=0.5 \times \frac{seqLen-digLen}{seqLen},
\end{equation}

\hspace{-3.2ex}where $seqLen$ is the log message length of the log event; $digLen$ is the number of tokens in the log event that contain digits. This is inspired by our observation that tokens containing digits are more likely to be variables. For each newly created log group in the DAG, its log event is the first log message in the group. Thus, following Equation \ref{equ:initst}, we can calculate an $st_{init}$ for each log group. Intuitively, $st_{init}$ estimates a lowerbound for the constant ratio in a log event. We consider the constant ratio in the similarity threshold design because if a log message belongs to a log group, it must have the same constants as the log event.

\subsection{Update an Existing Log Group}

\subsubsection{Update the DAG}
When Drain matches the current log message with an existing log group, Drain will add the log ID to the log ID list maintained by the log group. Besides, Drain will update the log event of the log group. Specifically, Drain scans the tokens in the same position of the log message and the log event. If the two tokens are the same, we do not modify the token in that token position. Otherwise, we update the token in that token position by wildcard (i.e., *) in the log event. For example, if the current log message ``Send file file\_01" matches an existing log group, whose log event is ``Send file file\_02", the updated log event will be ``Send file *".

\subsubsection{Similarity Threshold Updating}

The similarity threshold $st$ will update dynamically in the parsing process. When the log event of a log group is updated (i.e., one or more wildcards generated), Drain will update $st$ according to the number of updated tokens in the log event. Note that in existing log parsing methods, parameters (e.g., thresholds) is fixed manually for all log groups by developers beforehand. It is difficult for developers to find a threshold that works for all log groups. Drain not only provides customized thresholds (i.e., $st$) for each log group, but also updates the thresholds dynamically in the parsing process. Specifically, the threshold $st$ is updated by the following equation:

\begin{equation}
\label{equ:st}
st=\min \{1, st_{init}+0.5 \times \log_{base} (\eta+1)\},
\end{equation}

\hspace{-3.2ex}where $\eta$, which is initialzed to $0$ at the beginning, is the accumulated number of tokens that have been replaced by wildcards when updating the log events. In the parser runtime, a threshold $st$ is mainly changed with the value of $\eta$. When a token of the log event is replaced by ``wildcard" (leads to $\eta = \eta +1$), that token is regarded as a variable. Thus, the more variables found, the more difficult for a log message to get accepted to a log group. Intuitively, this is reasonable because we calculate the similarity between two sequences only based on constants, as illustrated in Equation \ref{equ:simseq}. At the beginning, the log event may contain some unfound variables, such as ``file\_01" in the initial log event ``Send file file\_01". If Drain sets a very strict similarity threshold, such as $1$ at the beginning, log messages that have the same log event (e.g., ``Send file file\_02") may not be able to get accepted to this log group. Thus, we gradually increase $st$ as more variables are found and replaced by wildcards. $base$ is defined as follows:

\begin{equation}
\label{equ:base}
base=\max\{2, digLen+1\},
\end{equation}

\hspace{-3.2ex}where $base$ and $digLen$ are both calculated when the log group is created, and will be fixed during the parsing process. In Equation \ref{equ:st}, $base$ controls the increasing speed of $st$. If $digLen$ is large, Drain thinks there may be many unfound variables, and thus will set a large base, which leads to a relatively slow increasing speed of $st$.

Although we cannot guarantee that the dynamic update mechanism is perfect for all system logs, we find it generally works well on a wide range of system logs, as demonstrated in our experiments.

\subsection{Merge Log Groups}\label{sec:mergeloggroup}

In pratice, developers may want to merge some over-parsed log groups. Thus, we provide an optional merge mechanism in the output layer of Drain. Specifically, if the merge mechanism is activated, Drain will check whether the new log event can be merged with an existing log event when a new log group is created.

Specifically, Drain provides a merge threshold $mt$ for developers to merge over-parsed log groups. When a new log group is created, Drain will compare the similarity between this log event and other existing log events, and return the most similar log event. If the similarity is larger than $mt$, Drain will merge the corresponding two log groups by pointing them to the same node in the output layer.

To calculate the similarity between two log events, we first find the longest common subsequence (LCS) of the two templates. For instance, sequence \{1,2,3,4\} and sequence \{2,4,5\} yield the LCS \{2,4\}. The similarity $temSim$ is defined as follows:

\begin{equation}
\label{equ:temSim}
temSim=\frac{lenLCS}{\min \{lenNew, lenExist\}},
\end{equation}

\hspace{-3.2ex}where $lenLCS$ is the number of tokens of the LCS; $lenNew$ is the number of tokens of the new log event; and $lenExist$ is the number of tokens of the existing log event. The LCS will also be regarded as the new log event for the merged groups. Note that although the merge threshold $mt$ is manually set, the merge mechanism is optional. For the experiments in this paper, we only use this mechanism in one log dataset (11 datasets in total).

\section{Datasets}\label{sec:dataset}

In the experiments, we mainly evaluated the performance of Drain in terms of accuracy and efficiency. However, one major limitation of the existing studies on log parsing \cite{SLCT03, ltang11, qfu09, IPLoM12, Mizutani13, spellICDM16, He16DSN, He17ICWS, HeTDSC17} is the lack of publicly released log datasets. For example, our previous conference version \cite{He17ICWS} employs five log datasets. Companies rarely release their log data to the public, because doing so may violate certain confidentiality clauses. To address this problem and facilitate research in this field, we collected 11 log datasets in this paper and released all of the 11 datasets online \cite{logparserdatasets}. This study is the first work that releases a large amount of datasets of great variety for log parsing tasks.

The log datasets used in our evaluation are summarized in Table \ref{tab:summaryofdatasets}. These 11 real-world datasets include distributed system logs (HDFS, Zookeeper, Hadoop, and Spark), Web application logs (Apache), supercomputer logs (BGL, HPC, and Thunderbird),  operating system logs (Windows and Linux), and standalone software logs (Proxifier). 

\begin{table}[t]
\protect\caption{Summary of Log Datasets\label{tab:summaryofdatasets}}
\includegraphics[scale=0.565]{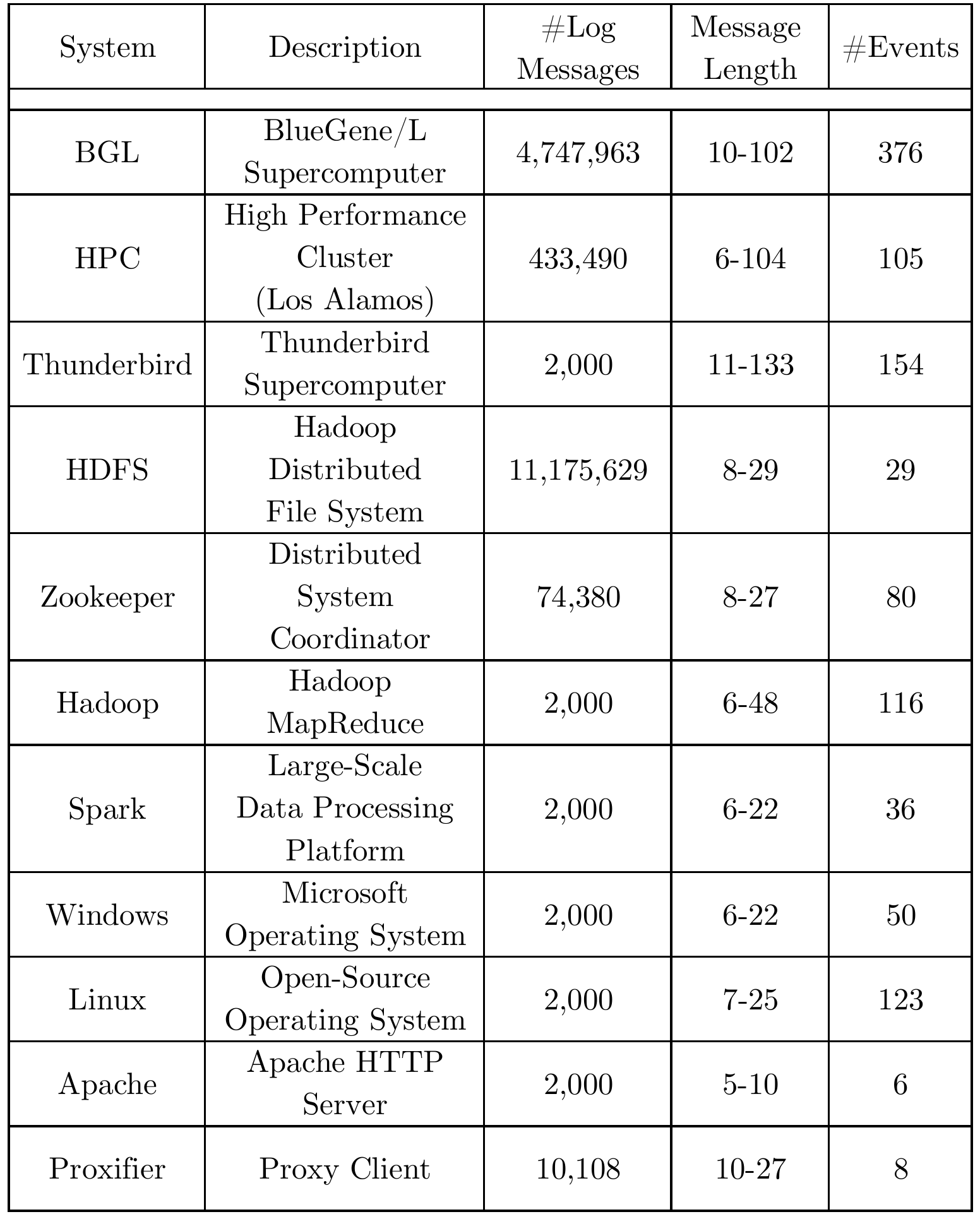}\center
\end{table}

We obtain 6 log datasets from other researchers with their generous support. Specifically, HDFS is a log dataset collected from a 203-node cluster on Amazon EC2 platform in \cite{weixu09}. Hadoop logs are collected by Microsoft from Hadoop MapReduce tasks \cite{linqwICSE16}. BGL is a log dataset collected by Lawrence Livermore National Labs (LLNL) from BlueGene/L supercomputer system \cite{BGLdata}. HPC logs are collected from a high performance cluster, which has 49 nodes with 6,152 cores and 128GB memory per node \cite{HPCdata}. Thunderbird is a log dataset collected from a supercomputer with 9,024 processors and 27,072GB memory in total \cite{BGLdata}. Apache and Linux are logs released by Chuvakin on a security log sharing site \cite{securitylogsharing}. We also collect 5 log datasets for evaluation. Zookeeper and Spark are datasets collected on a 32-node cluster in our laboratory. Windows and Proxifier are logs collected from laboratory computers running Windows 7.  As with the existing work \cite{ltang11, He16DSN}, each of the six extended datasets contains 2,000 log messages. We extract the 2,000 log messages by random extraction from the original messages. We conduct this sampling process because the original datasets are too large in size (e.g., 30GB for Thunderbird). 

After collecting all of the 11 log datasets, we spend a large effort to label them manually. In particular, for each log dataset, we create a list of log events by inspecting the raw log messages. Then, we use the log events to separate the raw log messages into different log groups, where the log messages in the same group shared identical log events. This labeling process is non-trivial, because there are a variety of log events that could only be found by manual inspection. For example, the dataset from Thunderbird contained 154 log events. Furthermore, each labeled dataset contains the following three parts: (1) the raw log messages, (2) a log event list, and (3) the separated log groups. All of the 11 datasets are released online \cite{logparserdatasets} for reproducible research.

\section{Evaluation}\label{sec:exper}

This section first describes the experimental settings. Then, we present an evaluation of the performance of Drain on the 11 considered datasets in terms of accuracy and efficiency. We further present a case study on the effectiveness of Drain in a system reliability management task (i.e., anomaly detection). Finally, we evaluate the automated parameter tuning mechanism of Drain.

\subsection{Experimental Settings}


\subsubsection{Comparison} To prove the effectiveness of Drain, we compare its performance with that of five existing log parsing methods. In particular, three of them are offline log parsers, and the other two are online log parsers. The ideas of these log parsers are briefly introduced as following:

\begin{itemize}
    \item LogSig \cite{ltang11}: LogSig is a clustering-based offline log parser. It provides novel distance calculation method based on token pairs. 
    \item LKE \cite{qfu09}: This is an offline log parsing method developed by Microsoft. It employs hierarchical clustering and heuristic rules. 
    \item IPLoM \cite{IPLoM12}: IPLoM conducts a three-step hierarchical partitioning before template generation in an offline manner.
    \item SHISO \cite{Mizutani13}: In this online parser, a tree with predefined number of children in each node is used to guide log group searching.
    \item Spell \cite{spellICDM16}: This method uses longest common sequence to search log group in an online manner. It accelerates the searching process by subsequence matching and prefix tree.
\end{itemize}


\subsubsection{Evaluation Metric and Experimental Setup} We apply F-measure \cite{clustering_evaluation}, which is a typical evaluation metric for clustering algorithms, to evaluate the accuracy of log parsing methods. The definition of accuracy is as following.

\begin{equation}
\label{equ:parsingaccuracy}
Parsing~Accuracy=\frac{2*Precision*Recall}{Precision+Recall},
\end{equation}

\hspace{-3.2ex}where $Precision$ and $Recall$ are defined as follows:

\begin{equation}
\label{equ:preciisonrecall}
Precision=\frac{TP}{TP+FP}, Recall=\frac{TP}{TP+FN},
\end{equation}


\hspace{-3.2ex}where a true positive ($TP$) decision assigns two log messages with the same log event to the same log group; a false positive ($FP$) decision assigns two log messages with different log events to the same log group; and a false negative ($FN$) decision assigns two log messages with the same log event to different log groups. This evaluation metric is also employed in our previous study \cite{He16DSN} on existing log parsers.

We run all experiments on a Linux server with Intel Xeon E5-2670v2 CPU and 128GB DDR3 1600 RAM, running 64-bit Ubuntu 14.04.2 with Linux kernel 3.16.0. We run each experiment 10 times to avoid bias. In the first layer (i.e., preprocessing layer) of Drain, we remove obvious parameters in log messages (i.e., IP addresses in HPC\&Zookeeper\&HDFS\&Thunderbird\&Linux, core IDs in BGL, block IDs in HDFS, and node IDs in HPC). Different from the previous conference version and the existing parsers, we do not need to tune parameters for our proposed parser Drain on most of the datasets. The only parameter used in our experiments is the merge threshold $mt$ (introduced in Section \ref{sec:mergeloggroup}) for the Proxifier dataset ($mt=0.95$). We apply this optional merge mechanism on Proxifier because its log messages were over-parsed in the length layer of Drain. For the other log parsers, we re-tune the parameters to optimize their performance.


\subsection{Accuracy of Drain}\label{sec:accuracy-of-drain}

\begin{table*}[t]
\centering
\protect\caption{Parsing Accuracy of Log Parsing Methods\label{tab:Parsing-Accuracy-of}}
\includegraphics[scale=1]{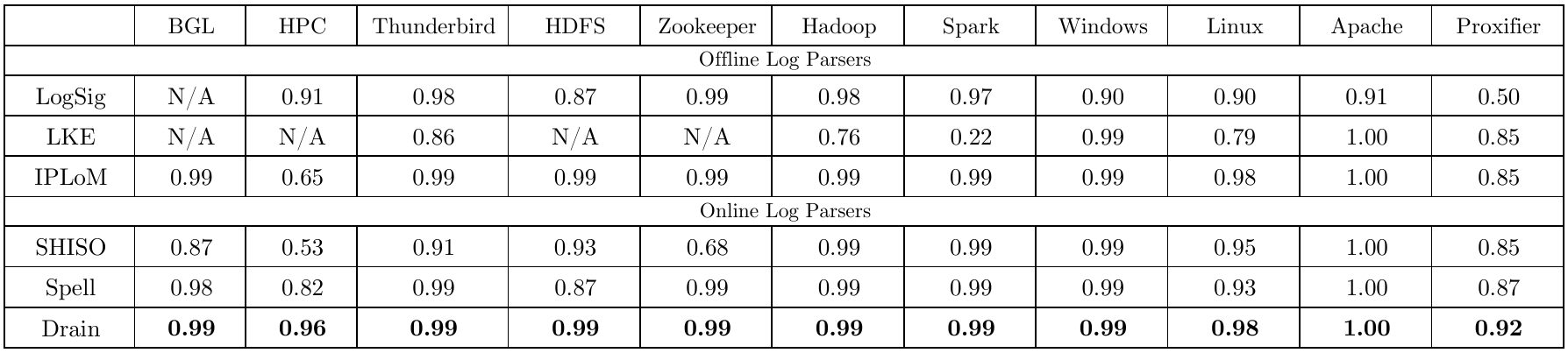}\center
\end{table*}

Accuracy demonstrates how well a log parser matches the raw log messages with the correct log events. Accuracy is important because parsing errors can degrade the performance of system reliability management tasks \cite{He16DSN}. Intuitively, an offline log parsing method can achieve higher accuracy than an online one, because an offline method obtains all of the raw log messages at the beginning of parsing, while an online method adjusts its parsing model gradually during the parsing process. 

In this section, we evaluate the accuracy of three offline and two online log parsing methods on the datasets described in Table \ref{tab:summaryofdatasets}. The evaluation results are in Table \ref{tab:Parsing-Accuracy-of}. LogSig and LKE cannot parse some datasets in reasonable time (e.g., hours or even days), whose results are marked as not available. LogSig fails to handle BGL, because each iteration of the underlying clustering algorithm takes too much time. LKE cannot handle BGL, HPC, HDFS, and Zookeeper, because its $O(n^2)$ time complexity with respect to the number of log messages makes it too slow for these datasets. 

We observe that the proposed online parsing method Drain obtains the best accuracy on all 11 datasets, even compared with the offline log parsing methods. LogSig does not perform well on Proxifier, because it requires the user to manually set the number of clusters (i.e., log groups), which leads to a trade-off between precision and recall. LKE is not that good on some datasets, because it employs an aggressive clustering strategy, which can lead to under-parsing. IPLoM obtains high accuracy on most datasets because of its specially designed heuristic rules. SHISO uses the similarity of characters in log messages to search the corresponding log events. This strategy is too coarse-grained, which causes inaccuracy. Spell is accurate, but its strategy only based on longest common subsequence can lead to under-parsing. Drain obtains 0.92$\sim$1.00 parsing accuracy, which is the highest compared with the existing log parsers on all of the 11 datasets. Drain achieves the highest accuracy because of the following reasons. First, it compounds both the log message length (length layer) and the first/last tokens of the log messages (token layer), which are effective, specially designed rules, to construct the DAG. Second, Drain only uses tokens that do not contain digits to guide the searching process (token layer), which effectively avoids over-parsing. Third, Drain provides an optional merge mechanism (output layer) that can handle over-parsing by the previous layers. Fourth, different from existing parsers, Drain can automatically initializes its parameters and updates them dynamically according to the incoming log messages. Last but not the least, Drain maintains different parameters for different log groups in runtime, while existing methods employ identical parameters for all the log groups.


\subsection{Efficiency of Drain}

In the following, we discuss the evaluation the efficiency of Drain. First, we compare the running time of Drain with that of the five existing log parsers on all of the 11 real-world datasets. Second, we evaluate the running time of these parsers on sample datasets of different sizes.

\subsubsection{Running Time on Real-World Datasets}

\begin{table*}[t]
\centering
\protect\caption{Running Time (Sec) of Log Parsing Methods\label{tab:runningtime}}
\includegraphics[scale=0.99]{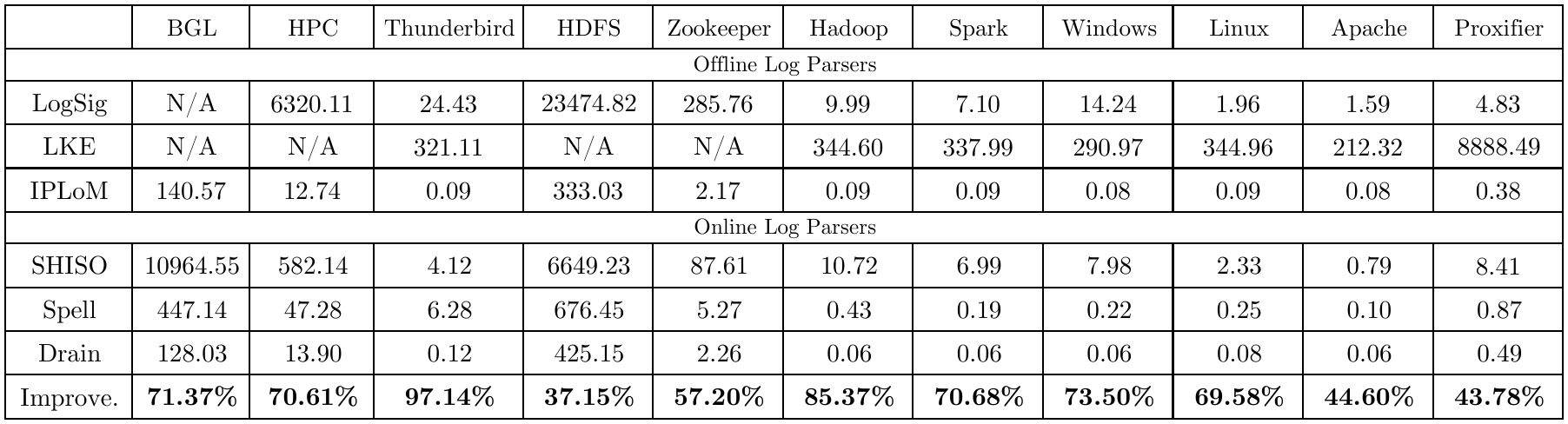}\center
\end{table*}

To evaluate the efficiency of Drain, we measure the running time of it and 5 existing log parsers on 11 real-world log datasets described in Table \ref{tab:summaryofdatasets}. In Table \ref{tab:runningtime}, we demonstrate the running time of these log parsers. Similar to the accuracy experiments in the previous section, LogSig and LKE fails to handle some datasets in reasonable time, so we mark the corresponding results as not available. Note that in the efficiency experiments, we mainly compare Drain with the online log parsers. The results of the offline parsers are presented for completeness. 

Considering online parsing methods, SHISO takes too much time on some datasets (e.g., takes more than 3h on BGL). This is mainly because SHISO only limits the number of children for its tree nodes, which can cause very deep parse tree. Spell obtains better efficiency performance, because it employs a prefix tree structure to store all log events found, which greatly reduces its running time. However, Spell does not restrict the depth of its prefix tree either, and it calculates the longest common subsequence between two log messages, which is time consuming. Compared with the existing online parsing methods, our proposed Drain requires the least running time on all 11 datasets. Specifically, Drain only needs 2 min to parse 4m BGL log messages and 7 min to parse 11m HDFS log messages. Drain greatly improves the running time of existing online parsing methods. The improvements on the 11 real-world datasets are at least $37.15\%$, and it reduce $97.14\%$ running time on Thunderbird. The efficiency improvement of Drain on these datasets are $65.54\%$ on average. 

Drain also demonstrates comparable efficiency with the state-of-the-art offline log parsing methods. It requires less running time than LogSig and LKE on all 11 datasets, and it takes similar running time compared with the state-of-the-art offline log parser IPLoM. Moreover, as an online log parsing method, Drain is not limited by the memory of a single computer, which is the bottleneck of most offline log parsing methods. For example, IPLoM needs to load all log messages into computer memory, and it will construct extra data structures of comparable size in runtime. Thus, although IPLoM is efficient too, it may fail to handle large-scale log data. Drain is not limited by the memory of single computer, because it processes the log messages one by one, not as a whole. 


\subsubsection{Running Time on Sample Datasets of Different Sizes}

\begin{table}[t]
\centering
\protect\caption{Log Size of Sample Datasets for Efficiency Experiments\label{tab:Parsing-Efficiency-of}}
\includegraphics[scale=0.64]{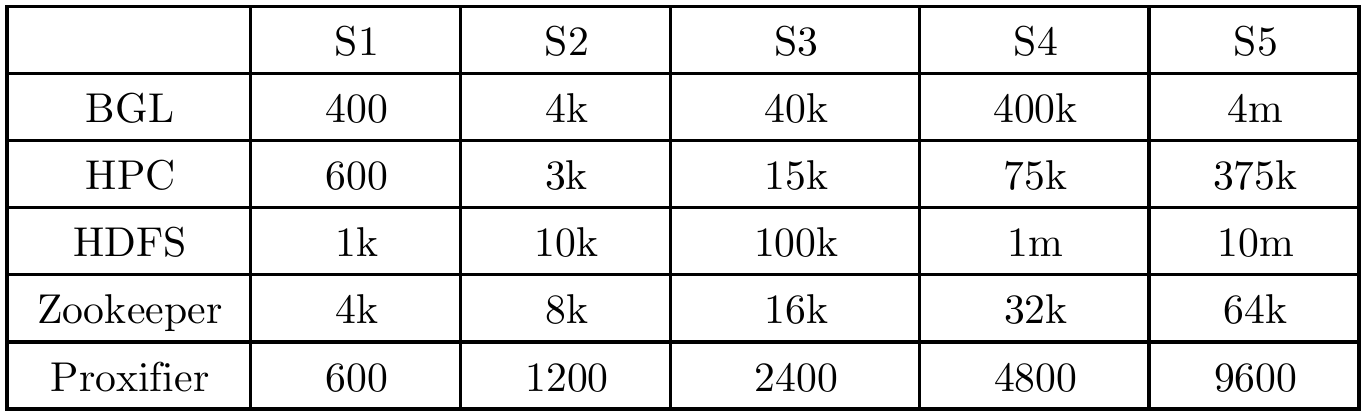}\center
\end{table}

Because log size of modern systems is rapidly increasing, a log parsing method is expected to handle large-scale log data. Thus, to simulate the increasing of log size, we also measure the running time of these log parsers on 25 sampled log datasets with varying log size (i.e., number of log messages) as described in Table \ref{tab:Parsing-Efficiency-of}. The log messages in these sampled datasets are randomly extracted from the real-world datasets in Table \ref{tab:summaryofdatasets}.

\begin{figure}
\centering{}
\vspace{0.5ex}
\begin{tabular}{c}
 \includegraphics[scale=0.94]{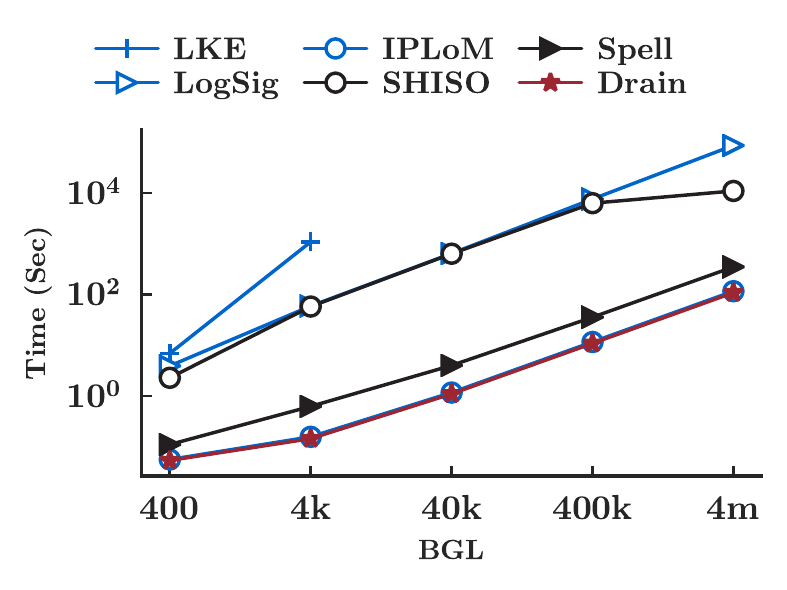}\vspace{-2ex} \\

 \includegraphics[scale=0.94]{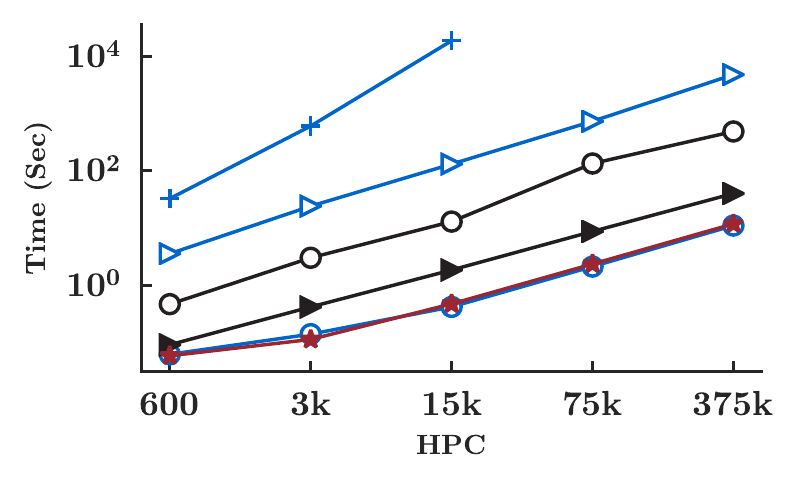}\vspace{-2ex} \\
 
 \includegraphics[scale=0.94]{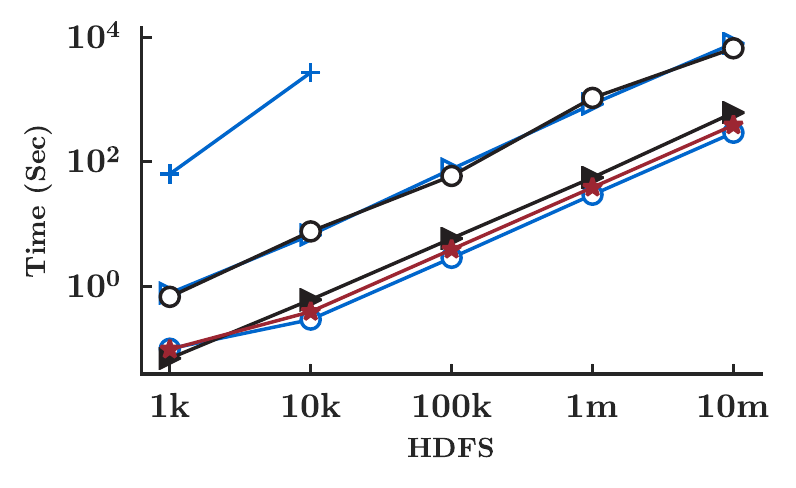}\vspace{-2ex} \\
 
 \includegraphics[scale=0.94]{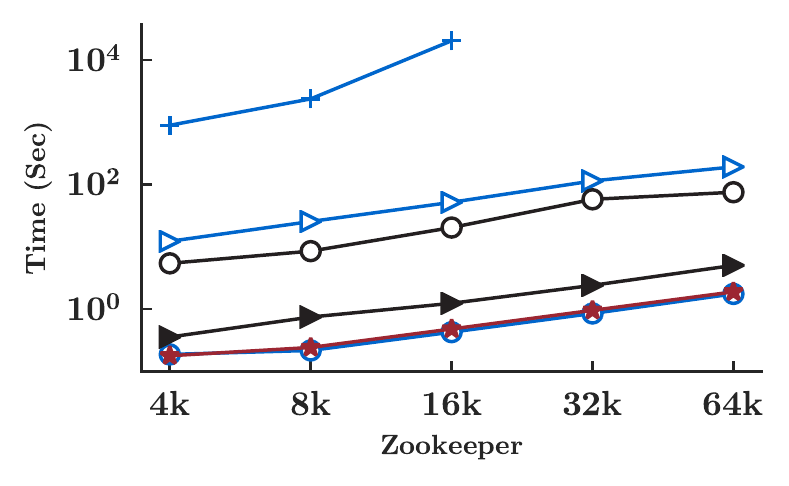}\vspace{-2ex} \\
 
 \includegraphics[scale=0.94]{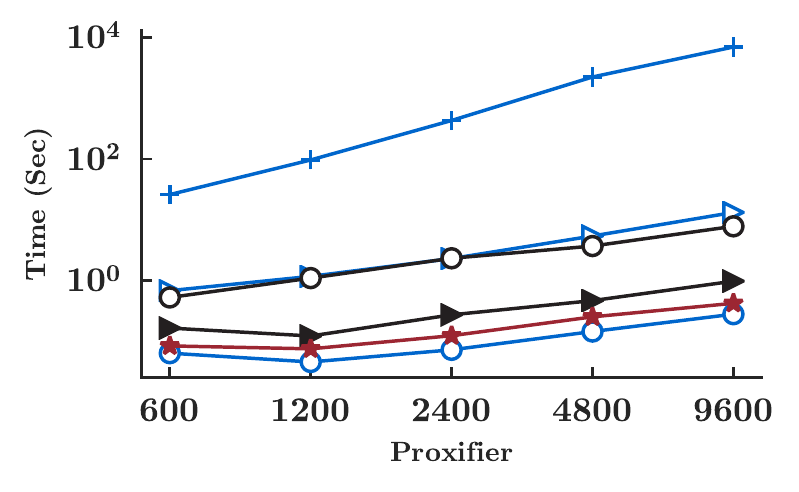} \\
 
\end{tabular}
\caption{Running Time of Log Parsing Methods on Datasets in Different Size}
\label{fig:running-time-log-parser}
\end{figure}

The evaluation results are illustrated in Fig. \ref{fig:running-time-log-parser}, which is in logarithmic scale. In this figure, we observe that, compared with other methods, the running time of LKE raises faster as the log size increases. Because the time complexity of LKE is $O(n^2)$, and the time complexity of other methods is $O(n)$, while $n$ is the number of log messages. Although the time complexity of LogSig is $O(n)$, it needs to generate token pairs for each log message, which causes an iteration of its clustering algorithm too slow. IPLoM is comparable to Drain, but it requires substantial amounts of memory as explained above. Online parsing methods (i.e., SHISO, Spell, Drain) process log message one by one, and they all use a data structure (i.e., a tree or a graph) to accelerate the log event search process. Drain is faster than the others because of three main reasons. First, Drain enjoys linear time complexity. The time complexity of Drain is $O(~(d + cm)n~)$, where $d$ is the depth of DAG, $c$ is the number of candidate log groups in the similarity layer nodes, $m$ is the log message length, and $n$ is the number of log messages. Obviously, $d$ and $m$ are constants. $c$ can also be regarded as a constant, because the quantity of candidate log groups in each similarity layer node is similar, and the number of log groups is far less than that of log messages. Thus, the time complexity of Drain is $O(n)$. For SHISO and Spell, the depth of the parse tree could increase during the parsing process. Second, we use the specially designed $simSeq$ to calculate the similarity between a log message and a log event. Its time complexity is $O(m_1 + m_2)$, while $m_1$ and $m_2$ are number of tokens of them respectively. In Drain, $m_1=m_2$. In comparison, SHISO and Spell calculate the longest common subsequence between two sequences, whose time complexity is $O(m_1 m_2)$. Third, Drain provides a cache mechanism that can memorize the previous search results in the length layer, which further reduces the running time.

\subsection{Effectiveness of Drain on Real-World Anomaly Detection Task}\label{sec:effectiveness}

We have demonstrated the superiority of Drain in terms of accuracy and efficiency in previous sections. Although high accuracy is necessary for log parsing methods, it does not guarantee good performance in system reliability management tasks. To evaluate the effectiveness of Drain in system reliability management tasks, we conduct a case study on a real-world anomaly detection task. Note that the goal of this case study is not to achieve better log mining results, but to validate Drain against other parsers. We want to answer the following question: Even though the parsing accuracy of Drain is the highest, is it also effective in a system reliability management task?

In this case study, we use the HDFS log dataset. Specifically, raw log messages in the HDFS dataset \cite{weixu09} record system operations on 575,061 HDFS blocks with a total of 29 log event types. Among these blocks, 16,838 are manually labeled as anomalies by the original authors. In the original paper \cite{weixu09}, the authors use Principal Component Analysis (PCA) to detect these anomalies. Next, we will briefly introduce the anomaly detection workflow, including log parsing and log mining. In log parsing step, all the raw log messages are parsed into structured log messages. In log mining, we first use the structured log messages to generate an event count matrix, where each row represents an HDFS block; each column represents a log event type; each cell counts the occurrence of an event on a certain HDFS block. Then we use TF-IDF \cite{TFIDF87} to preprocess the event count matrix. Finally, the event count matrix is fed into PCA for anomaly detection.

In our case study, we evaluate the performance of the anomaly detection task with different log parsing methods used in the parsing step. Specifically, we use different log parsing methods to parse the HDFS raw log messages respectively and, hence, we obtain different sets of structured log messages. For example, an HDFS block ID could match with different log events by using different log parsing methods. Then, we generate different event count matrices, and feed them into PCA, respectively.

\begin{table}
\protect\caption{Anomaly Detection with Different Log Parsing Methods (16,838 True Anomalies) \label{tab:anomalydetection}}
\includegraphics[scale=0.54]{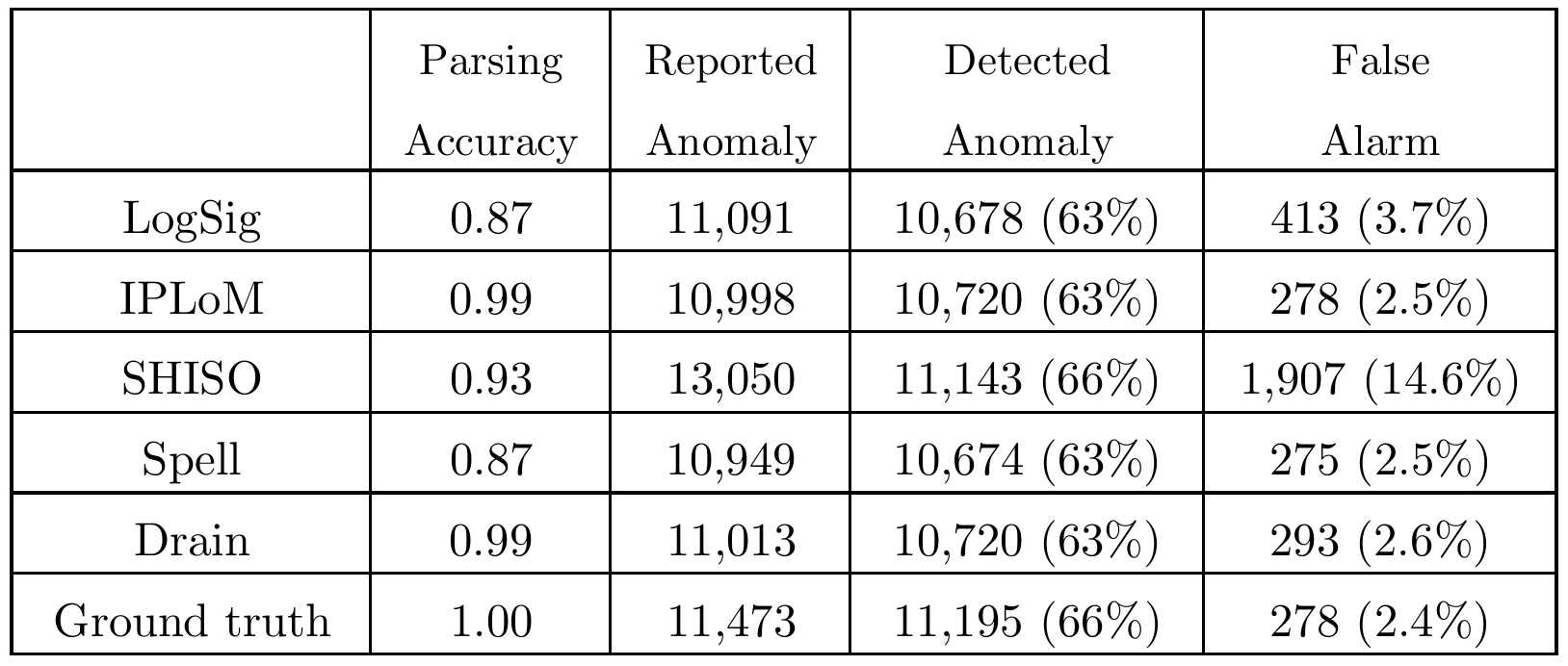}\center
\vspace{-4ex}
\end{table}

The experimental results are shown in Table \ref{tab:anomalydetection}. In this table, $reported$ $anomaly$ is the number of anomalies reported by the PCA model; $detected$ $anomaly$ is the number of true anomalies reported; $false$ $alarm$ is the number of wrongly reported ones. We use five log parsing methods to handle the parsing step of this anomaly detection task. We do not use LKE because it cannot handle this large amount of data. $Ground$ $truth$ is the experiment using exactly correct parsed results. 

We can observe that Drain obtains nearly the optimal anomaly detection performance. It detects $10,720$ true anomalies with only $293$ false alarms. Although $37\%$ of anomalies have not been detected, we argue that it is mainly caused by the log mining step. Because even when all the log messages are correctly parsed, the log mining model still leaves $34\%$ of anomalies at large. We also observe that SHISO, although has a high parsing accuracy (0.93), does not perform well in this anomaly detection task. By using SHISO, we would report $1,907$ false alarms, which are 6 times worse than others. This will largely increase the workload of developers, because they usually need to manually check the anomalies reported. Spell has nearly optimal anomaly detection performance because the model PCA is robust against its parsing errors. In this special case, Drain is comparable to Spell, with 18 more false alarms yet 46 more detected anomalies. Among the online parsing methods, Drain not only has the highest parsing accuracy as demonstrated in Section \ref{sec:accuracy-of-drain}, but also obtains nearly optimal performance in the anomaly detection case study.

\subsection{Effect of Automated Parameter Tuning}

\begin{figure}
\centering{}
 \includegraphics[scale=0.5]{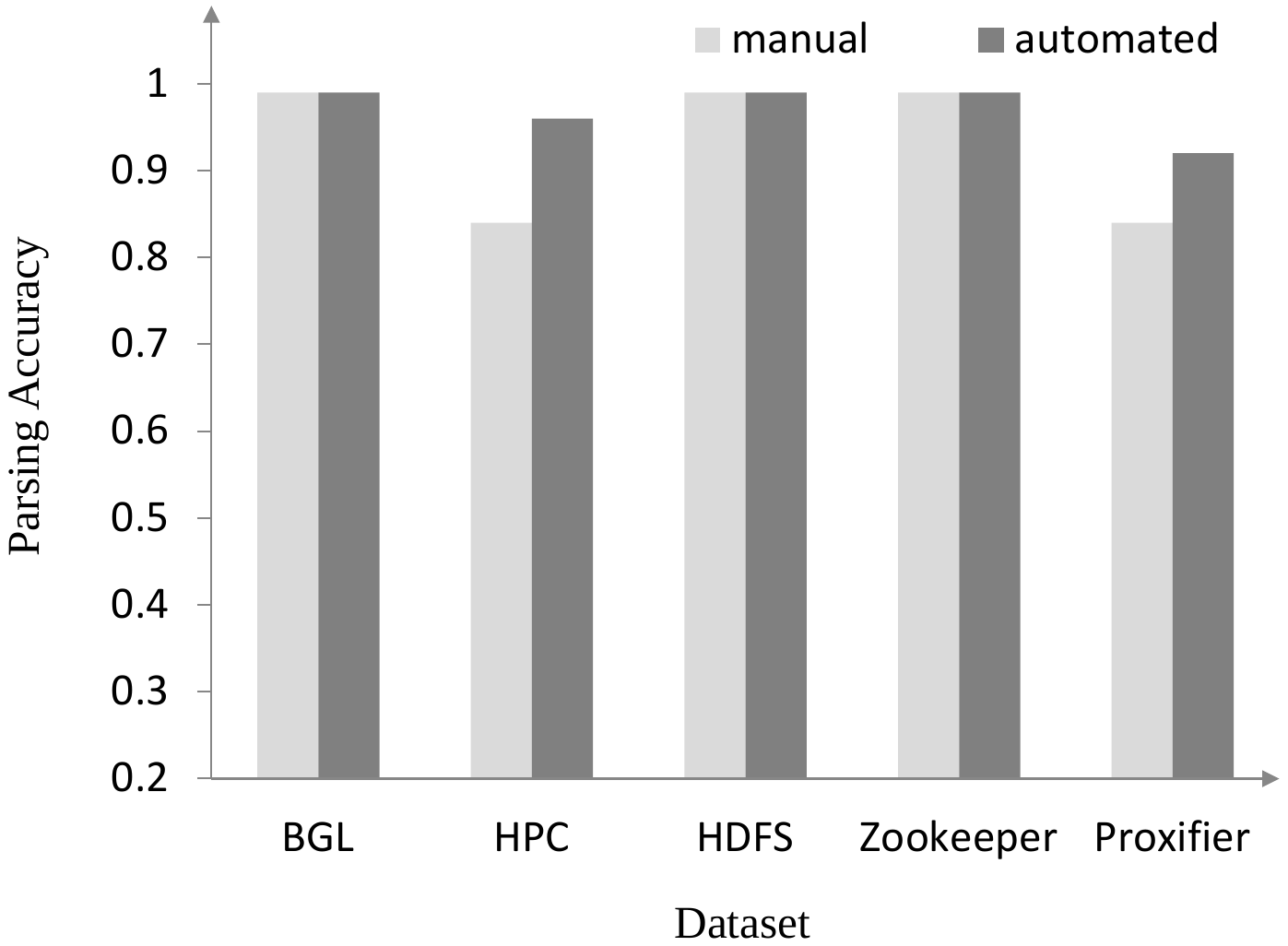}  
 
\caption{Effect of Automated Parameter Tuning}
\label{fig:para}
\end{figure}

\begin{figure}
\centering{}
 \includegraphics[scale=0.5]{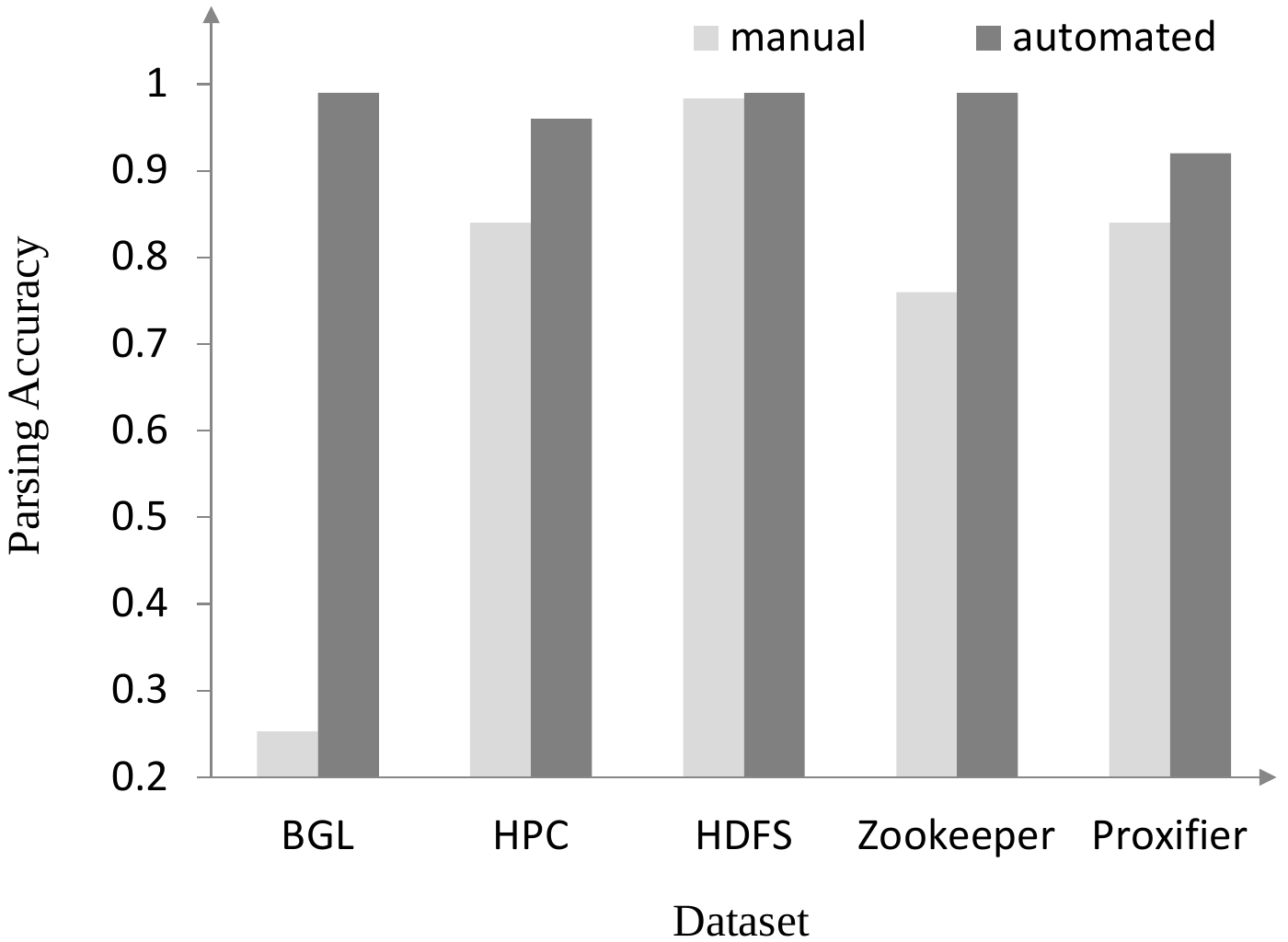}  
 
\caption{Apply Parameters Tuned on Proxifier for the Manual Parameter Tuning Method}
\label{fig:parahpc}
\end{figure}

In this section, we intend to evaluate the effectiveness of the automated parameter tuning mechanism of Drain, as presented in Section \ref{sec:update}. In our previous conference version \cite{He17ICWS}, we require developers to set two parameters: \textit{st} and \textit{depth}. In Fig. \ref{fig:para}, we compare the parsing accuracy of manually tuned parser \cite{He17ICWS} and the automated parameter tuning method Drain. The results show that, with the automated parameter tuning mechanism, Drain achieves comparable accuracy on BGL, HDFS, and Zookeeper, and outperforms the manual method on HPC and Proxifier. This demonstrates the effectiveness our the automated parameter tuning mechanism. Moreover, in Fig. \ref{fig:parahpc}, for the manual method, we directly apply the parameters tuned on the Proxifier dataset, and compare its accuracy with Drain. As illustrated, Drain achieves comparable or higher parsing accuracy. This shows that for the manual method, parameters tuned on a system cannot be directly used by other systems, while Drain consistently achieves high accuracy.

 \section{Discussions}\label{sec:dis}

In this section, we discuss the limitations of this work and provide some potential directions for future exploration.

\textbf{Diversity of system reliability management tasks.} In the experiment part, we conduct a case study on the effectiveness of log parsing methods in a system reliability management task (i.e., anomaly detection). We do not run experiments on more case studies because log datasets of reliability management tasks are scarce. Industrial companies are reluctant to release their log datasets due to privacy concerns. However, the anomaly detection task evaluated is an important and widely adopted task in both research and practice, which is presented in a paper with more than 400 citations. The effectiveness of our proposed parser Drain is demonstrated in this anomaly detection task. We consider collecting more log datasets used in reliability management tasks of distributed system as our future work.

\textbf{Logging of event ID.} Adding log event IDs into the logging statements can improve the accuracy of log parsing, because developers can generate structured logs based on log event IDs directly. Besides, developers who design the logging statement know exactly the corresponding  log event. Thus, adding log event IDs into the logging statements is a good logging practice from the perspective of automated log analysis \cite{logWritingFormat04}. In the future, we will explore automated log event ID adding methods, which can enhance the logging statements of modern systems.

\textbf{Distributed online parser.} Nowadays, a log management system typically has a log shipper installed on each node to forward log messages to a centralized server for further parsing \cite{spellICDM16}. The parser proposed in this paper aligns well with this mechanism in practice. In the future, we consider to conduct both log collection and log parsing in a distributed manner. Specifically, a parser will parse the collected logs in each node after log collection, and send the structured logs to the centralized server. A distributed online parser can accelerate the parsing process by parallelization and improve the efficiency of log management by reducing the size of transmitted log data.

\section{Related Work}\label{sec:related}

\textbf{Log parsing.}
Log parsing has been widely studied in recent years. Xu et al. \cite{weixu09} design a source code based log parser that achieves high accuracy. However, source code is often inaccessible in practice (e.g., third-party libraries or Web services). Some other work studies data-driven approaches, in which data mining techniques or heuristic rules are employed to extract log events and split raw log messages into different log groups accordingly. There are two kinds of data-driven log parsers: offline log parsers \cite{SLCT03,ltang11,qfu09,IPLoM12,HeTDSC17} and online log parsers \cite{Mizutani13,spellICDM16,He17ICWS}. Among the offline log parsers, SLCT \cite{SLCT03} and IPLoM \cite{IPLoM12} are based on heuristic rules summarized from the characteristics of system logs. LogSig \cite{ltang11} and LKE \cite{qfu09} employ clustering algorithms with specially designed distance metrics. These four log parsers have been studied in our prior work \cite{He16DSN} and our implementation has been released for future reuse. POP \cite{HeTDSC17} is a parallel offline parser that uses both heuristic rules and clustering algorithm. The goal of POP is to accelerate the offline parsing process by parallelization. Thus, we do not compare our online parser with it. SHISO \cite{Mizutani13} and Spell \cite{spellICDM16} are online log parsers, which parse log messages in a streaming manner, and are not limited by the memory of a single computer. Both of them design a tree structure to accelerate the log group search process. In this paper, we propose an online log parser, namely Drain, that greatly outperforms existing online log parsers in terms of both accuracy and efficiency. Different from existing offline parsers and online parsers, Drain can initialize its parameters automatically and update them dynamically in runtime, which further releases the burden of developers.

\textbf{Log analysis.} To enhance the reliability of modern distributed systems, various log analysis methods have been proposed in recent years. Xu et al. \cite{weixu09} design an anomaly detection method based on principal component analysis (PCA). Fu et al. \cite{qfu09} detect anomalies in distributed systems by learning a finite state machine from system logs. Kc et al. \cite{helenGuSRDS11} proposed a hybrid anomaly detection method employing both coarse-grained clustering and message flow graph mining. As for performance problem diagnosis, Nagaraj et al. \cite{Nagaraj12} propose a structured comparative analysis method. Log analysis is also employed in program verification. Beschastnikh et al. \cite{Beschastnikh11} design a tool called Synoptic, which builds a finite state machine based on system logs. Shang et al. \cite{wshang13} compare log sequences generated in lab testing environment and in large-scale cloud environment, which assists developers in detecting deployment bugs. Most of these log analysis methods require log parsing as their first step, which transform the unstructured logs into structured events. In this paper, we propose an online log parsing method that achieves the state-of-the-art performance in terms of both parsing accuracy and efficiency.

\textbf{Log management.}
Log management has become a challenging problem as distributed systems are large-scale in size and complex in structure. Besides the log parsing and log analysis methods introduced, there are various log management solutions available, such as commercial Splunk \cite{splunk}, and open-source solution ELK (i.e., Elasticsearch \cite{elasticsearch}, Logstash \cite{logstash}, Kibana \cite{kibana}). These log management solutions focus on the collection, indexing, searching, filtering, analyzing, and visualization of system logs. However, log parsing in these existing solutions require developers to manually construct matching patterns (e.g., regular expressions). In this paper, we propose an online log parsing method, which can automatically construct matching patters and dynamically update them in a streaming manner. Thus, we think the proposed online parser and existing log management solutions complement with each other.

\section{Conclusion}\label{sec:con}

This paper targets online log parsing for modern systems, particularly distributed systems. We propose an online log parser, called Drain, which can parse log messages in a streaming manner and update its parsing rules dynamically. Drain uses a DAG to encode specially designed heuristic rules. Different from the existing log parsers, Drain automatically initializes its parameters according to the characteristics of the incoming log messages and dynamically updates its parameters during runtime. Thus, Drain does not require manual parameter tuning efforts in most cases. Furthermore, to evaluate the performance of Drain, we collected 11 log datasets from real-world systems, ranging from distributed systems, Web applications, supercomputers, operating systems, to standalone software. Drain achieves the highest parsing accuracy on all of the 11 datasets. Meanwhile, Drain obtains 37.15\%$\sim$ 97.14\% improvement in the running time over the state-of-the-art online log parsers. The source code of Drain and all of the considered datasets have been publicly released to make them reusable and thus facilitate future research.

%
%
%
%




\ifCLASSOPTIONcaptionsoff
  \newpage
\fi

\bibliographystyle{IEEEtran}
\bibliography{References}


%

\vspace{-11ex}
\begin{IEEEbiography}[{\includegraphics[width=1in,height=1.25in,clip,keepaspectratio]{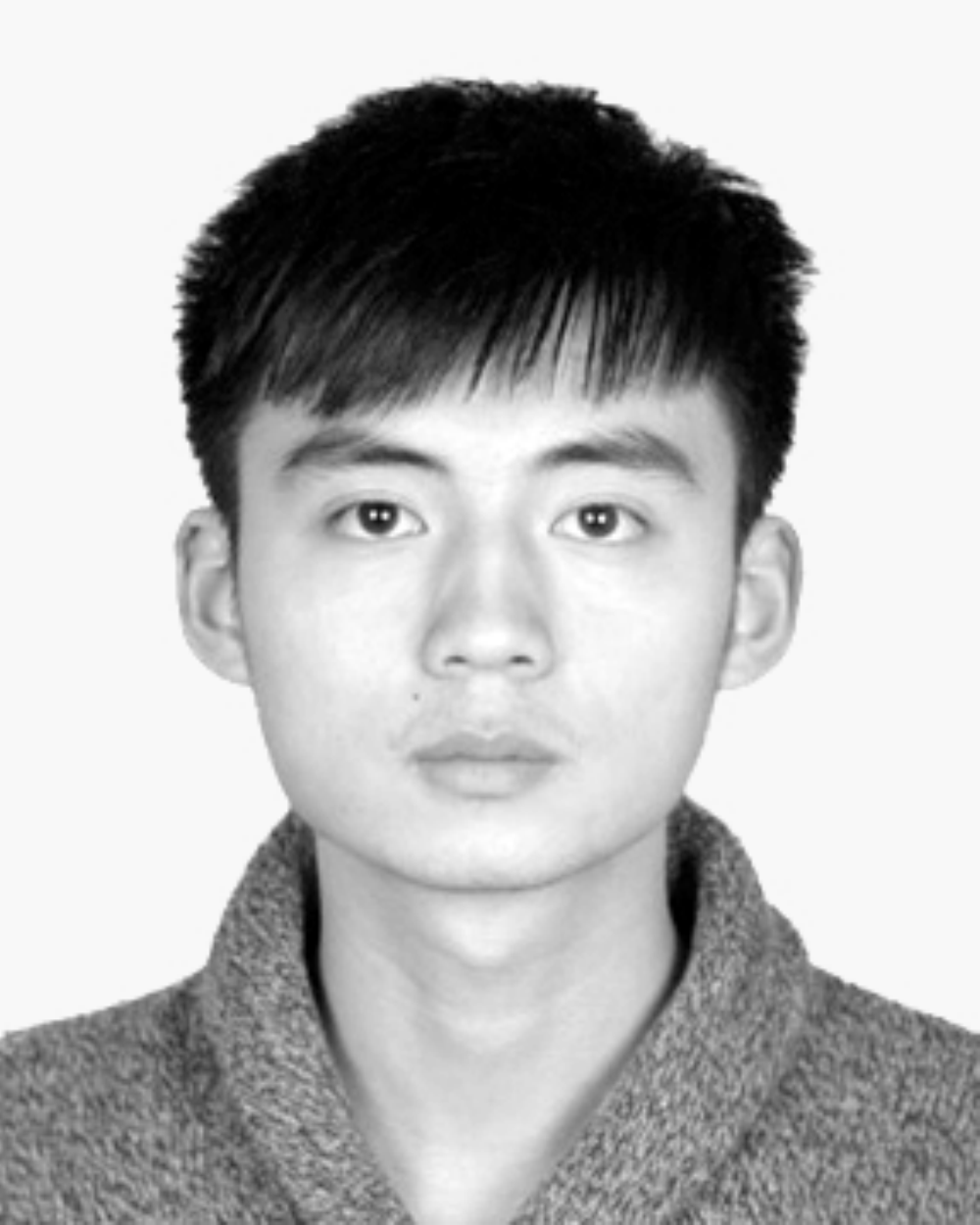}}]{Pinjia He}
received the BEng degree in Computer Science and Technology from South China University of Technology, Guangzhou, China. He is currently working towards the PhD degree in computer science and engineering in The Chinese University of Hong Kong, Hong Kong. His current research interests include log analysis, system reliability, software engineering and distributed system. He had served as an external reviewer in many top-tier conferences.
\vspace{-12ex}
\end{IEEEbiography}

\begin{IEEEbiography}[{\includegraphics[width=1in,height=1.25in,clip,keepaspectratio]{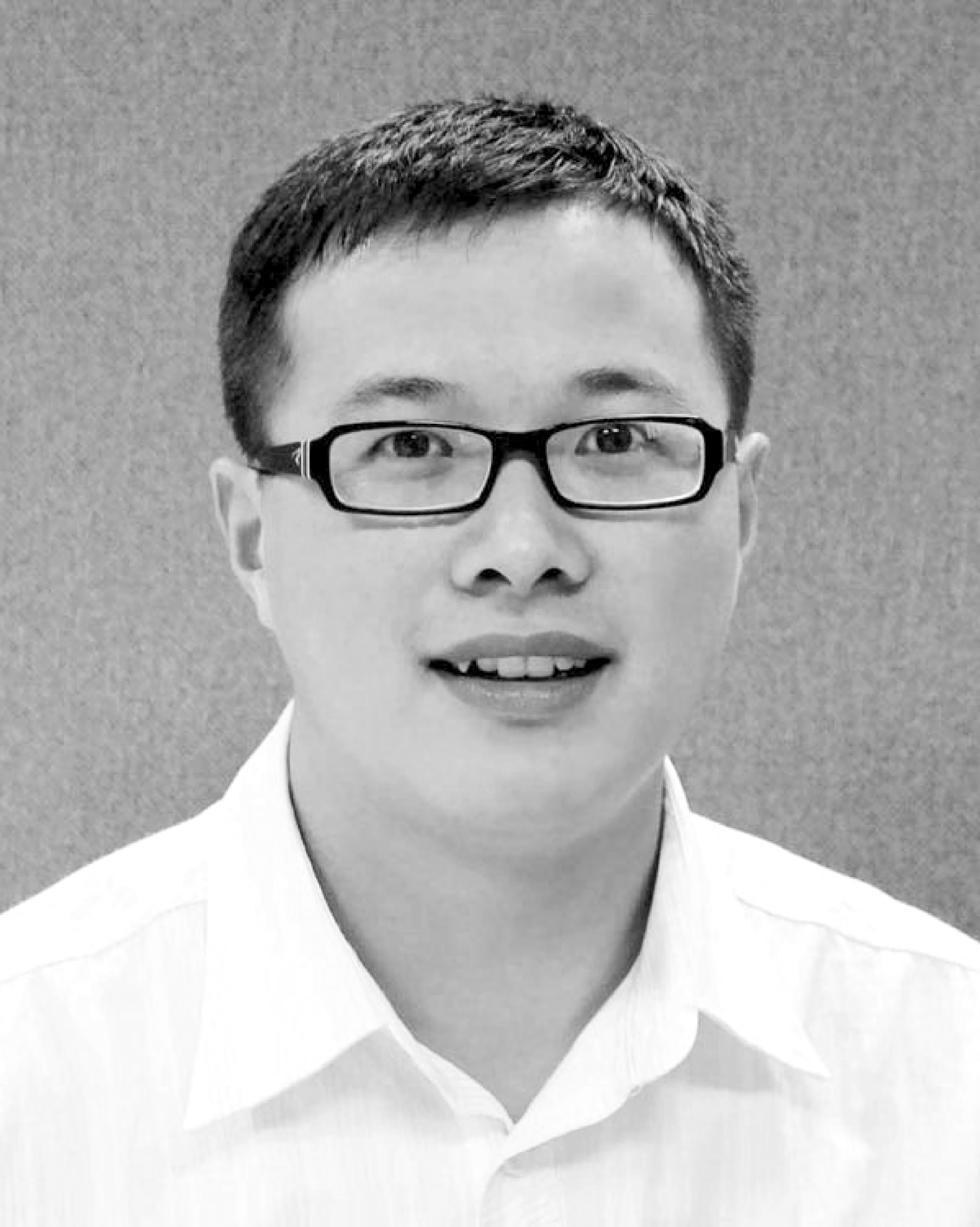}}]{Jieming Zhu}
is currently a postdoctoral fellow at The Chinese University of Hong Kong. He received the B.Eng. degree in information engineering from Beijing University of Posts and Telecommunications, Beijing, China, in 2011; the Ph.D. degree from Department of Computer Science and Engineering, The Chinese University of Hong Kong, in 2016. He served as an external reviewer for international conferences and journals including TSC, ICSE, FSE, WWW, KDD, AAAI, SRDS, ISSRE, ICWS, etc. His current research focuses on data monitoring and analytics for system intelligence.
\vspace{-12ex}
\end{IEEEbiography}

\begin{IEEEbiography}[{\includegraphics[width=1in,height=1.25in,clip,keepaspectratio]{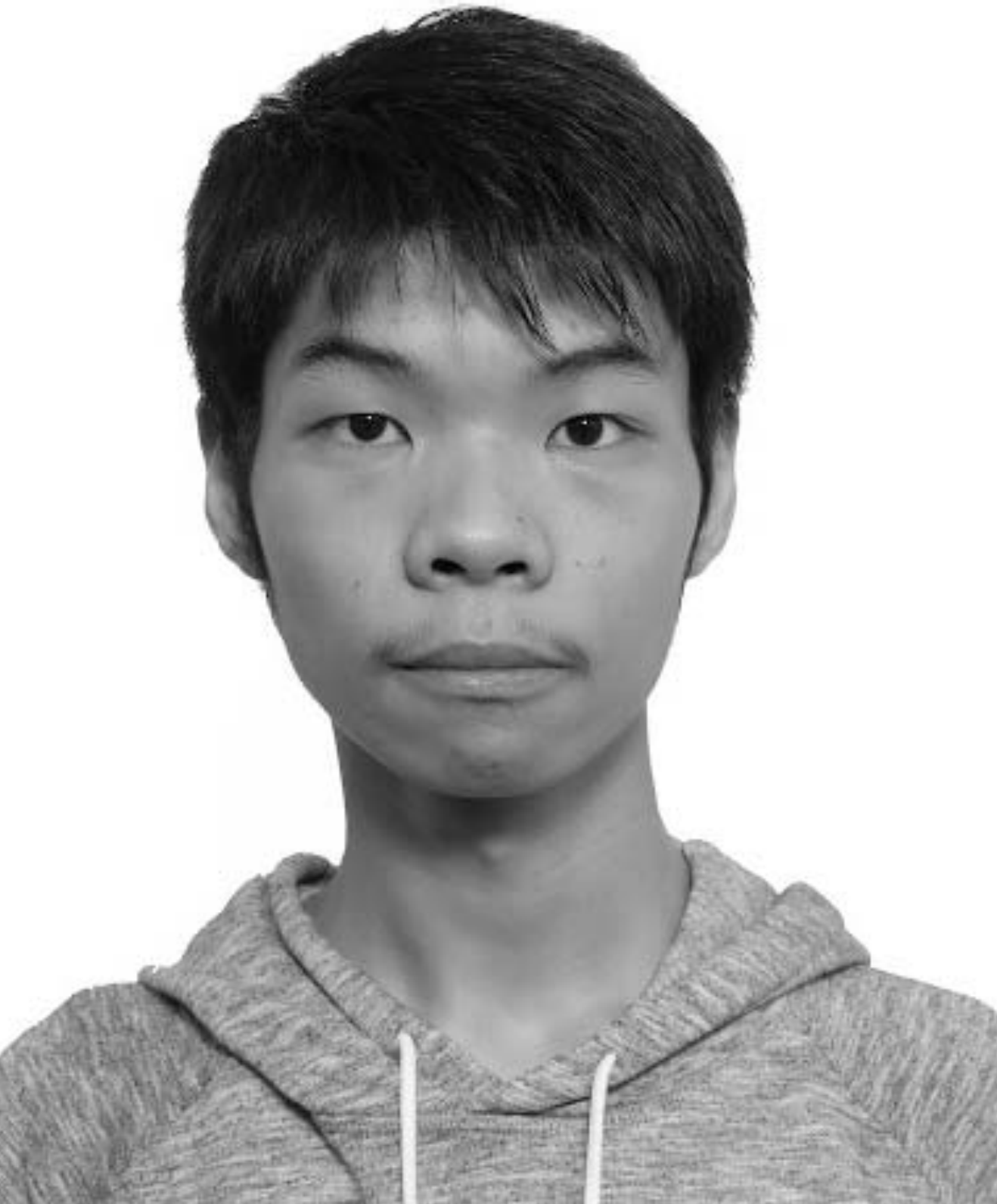}}]{Pengcheng Xu}
Pengcheng Xu is currently working towards the BSc degree in Computer Science in the Chinese University of Hong Kong, Hong Kong. His current research interests is log anglysis.
\vspace{-12ex}
\end{IEEEbiography}

\begin{IEEEbiography}[{\includegraphics[width=1.1in,height=1.375in,clip,keepaspectratio]{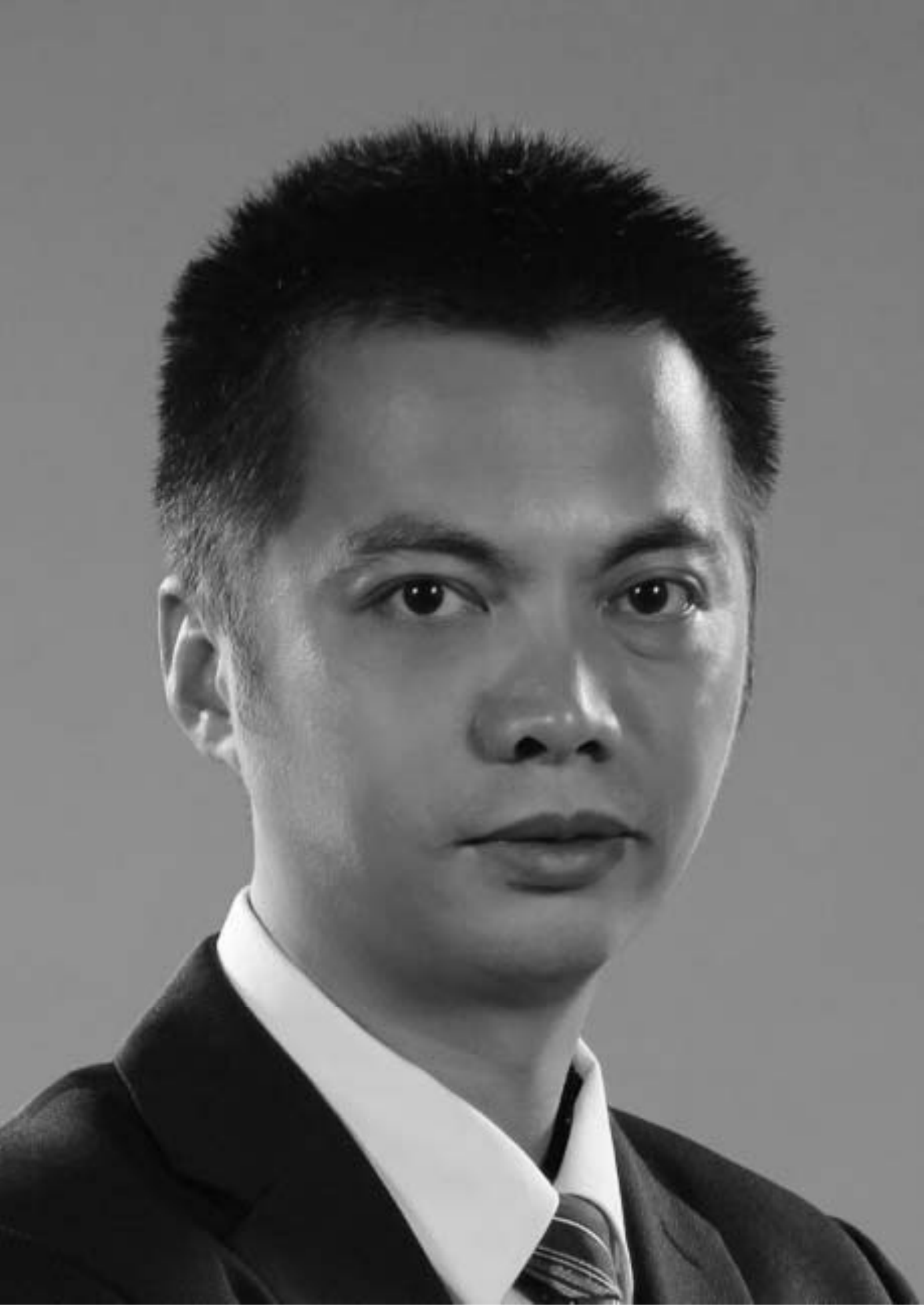}}]{Zibin Zheng}
is currently an associate professor at Sun Yat-sen University, Guangzhou, China. He received his Ph.D. degree from Department of Computer Science and Engineering, The Chi- nese University of Hong Kong, Hong Kong, in 2010. He received Outstanding Thesis Award of CUHK at 2012, ACM SIGSOFT Distinguished Paper Award at ICSE2010, Best Student Paper Award at ICWS2010, and IBM Ph.D. Fellowship Award 2010-2011. He served as program com- mittee member of IEEE CLOUD2009, SCC2011, SCC2012, ICSOC2012, etc. His research interests include cloud com- puting, service computing, and software engineering.
\vspace{-12ex}
\end{IEEEbiography}

\begin{IEEEbiography}[{\includegraphics[width=1in,height=1.25in,clip,keepaspectratio]{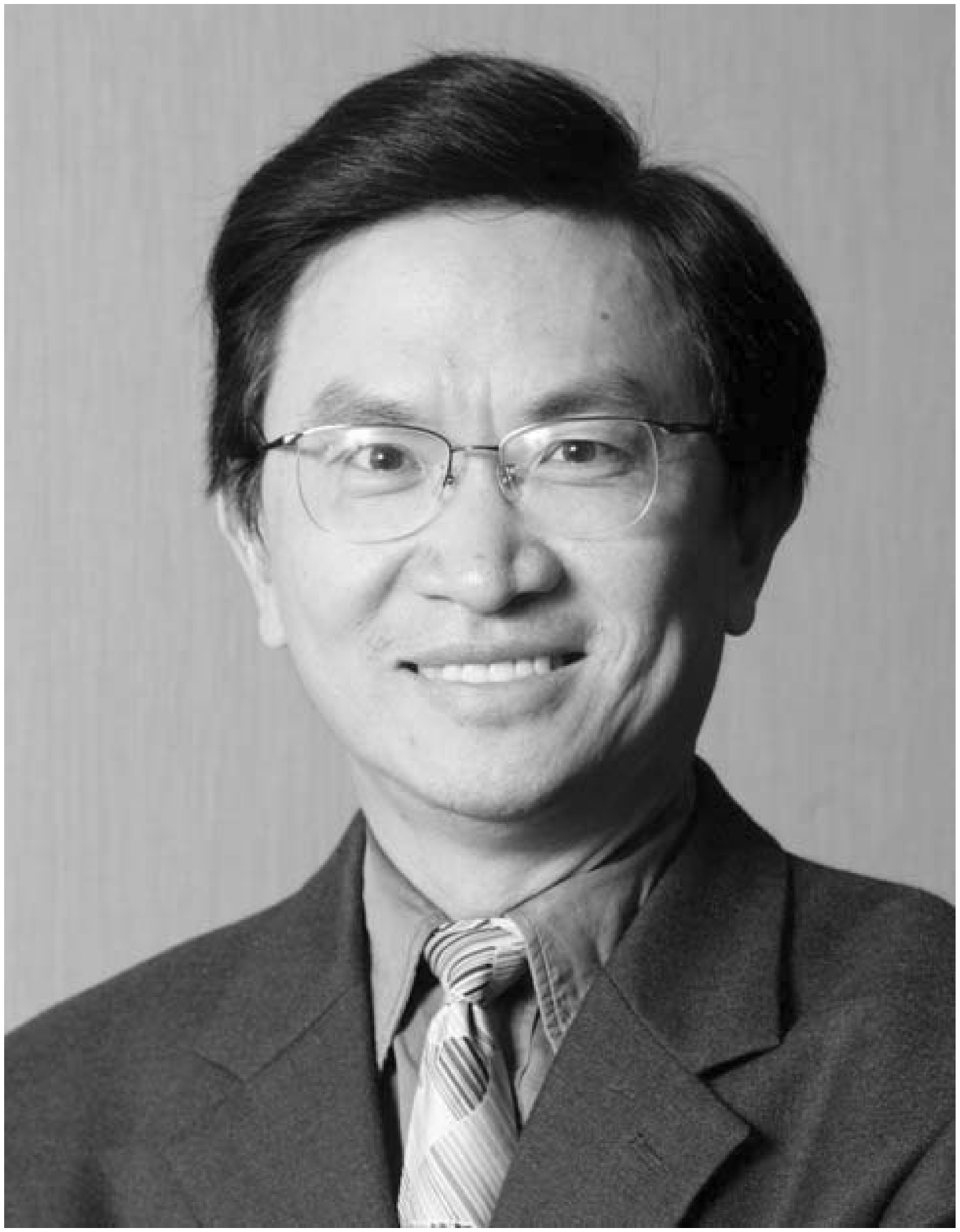}}]{Michael R. Lyu}
is currently a professor of Department of Computer Science and Engineering, The Chinese University of Hong Kong. He received the B.S. degree in electrical engineering from National Taiwan University, Taipei, Taiwan, R.O.C., in 1981; the M.S. degree in computer engineering from University of California, Santa Barbara, in 1985; and the Ph.D. degree in computer science from the University of California, Los Angeles, in 1988. His research interests include software reliability engineering, distributed systems, fault-tolerant computing, and machine learning. Dr. Lyu is an ACM Fellow, an IEEE Fellow, an AAAS Fellow, and a Croucher Senior Research Fellow for his contributions to software reliability engineering and software fault tolerance. He received IEEE Reliability Society 2010 Engineering of the Year Award.
\vspace{-12ex}
\end{IEEEbiography}




\end{document}